\documentclass[12pt,epsfig,epsf]{article}

\usepackage{cite}
\usepackage{slashed}
\usepackage{epsf,color,colordvi}
\usepackage{graphicx}
\usepackage{amsmath}
\usepackage{amssymb}
\usepackage{epsfig}
\usepackage{cite}
\usepackage{fontenc}
\usepackage{float}
\usepackage[lofdepth,lotdepth,caption=false]{subfig}
\usepackage{color}
\usepackage{booktabs}
\usepackage{tabularx}
\usepackage{multirow}
\usepackage{longtable}
\usepackage{lscape}
\usepackage{dcolumn}
\usepackage{rotating}
\usepackage[english]{babel}
\usepackage[autostyle]{csquotes}
\usepackage{hyperref}
\usepackage{xcolor}
\usepackage[normalem]{ulem}

\newcommand{\bi}{\begin{itemize}}
\newcommand{\ei}{\end{itemize}}

\def\beq{\begin{equation}}
\def\eeq{\end{equation}}

\newcommand{\bea}{\begin{eqnarray}}
\newcommand{\eea}{\end{eqnarray}}

\newcommand{\beqa}{\begin{eqnarray}}
\newcommand{\eeqa}{\end{eqnarray}}

\newcommand{\ldm}{\Delta m_{31}^2}
\newcommand{\sdm}{\Delta m_{21}^2}

\newcommand{\ie}{{\it i.e.}}

\newcommand{\eg}{{\it e.g.}}

\newcommand{\eq}{Eq.}
\newcommand{\eqs}{Eqs.}

\def\lsim{\mathrel{\rlap{\lower4pt\hbox{\hskip1pt$\sim$}}
    \raise1pt\hbox{$<$}}}         
\def\gsim{\mathrel{\rlap{\lower4pt\hbox{\hskip1pt$\sim$}}
    \raise1pt\hbox{$>$}}}         
\newcommand{\re}[1]{\ensuremath{{\cal R}e\left(#1\right)}}
\newcommand{\im}[1]{\ensuremath{{\cal I}m\left(#1\right)}}

\newcommand{\ham}{\mathcal{H}}

\DeclareMathOperator{\diag}{diag}

\newcommand{\equ}[1]{\eq~\ref{equ:#1}}

\newcommand{\equs}[2]{\eqs~\ref{equ:#1} and \ref{equ:#2}}

\def\ket#1{| \,#1\, \rangle}
\def\bra#1{\langle \,#1\, |}

\begin{document}

\begin{titlepage}

\renewcommand{\thefootnote}{\alph{footnote}}

\vspace*{1.cm}

\begin{flushright}
\end{flushright}

\renewcommand{\thefootnote}{\fnsymbol{footnote}}
\setcounter{footnote}{0}

\renewcommand{\thefootnote}{\fnsymbol{footnote}}
\setcounter{footnote}{-1}

{\begin{center}
{\large\bf 
 Investigating Leggett-Garg inequality in neutrino oscillations - role of  decoherence and decay
 \\
%
%
[0.2cm]
}
\end{center}}

\renewcommand{\thefootnote}{\alph{footnote}}

\vspace*{.8cm}
\vspace*{.3cm}
{
\begin{center} 
            {\sf 
                Sheeba Shafaq$^\S$\,\footnote[1]
               {\makebox[1.cm]{Email:} sheebakhawaja7@gmail.com}, 
                Tanmay Kushwaha$^\S$\,\footnote[2]
               {\makebox[1.cm]{Email:} tanmay.2811@gmail.com} 
               and 
                Poonam Mehta$^{\S}$\,\footnote[3]{\makebox[1.cm]{Email:}
                pm@jnu.ac.in}
} 
\end{center}
}
\vspace*{0cm}
{\it 
                                                                                               \begin{center}
$^\S$\, School of Physical Sciences,   \\   Jawaharlal Nehru University, 
      New Delhi 110067, India  \\

\end{center}}

\vspace*{1.5cm}

\begin{center}
{\Large \today}
\end{center}

{\Large 
\bf
 \begin{center} Abstract  
\end{center} 
}

Neutrinos, due to their weakly interacting nature, provide us with a unique opportunity to test foundations of quantum mechanics over macroscopic distances.  There has been considerable theoretical and experimental  interest in examining the extent of violation of Leggett-Garg inequalities (LGI) in the context of  neutrino flavour oscillations. 
While it is established that neutrino oscillations occur due to  mass and mixing among the three generations of neutrinos,  sub-dominant effects due to physics beyond the Standard Model (SM) are not yet ruled out. As we are entering the precision era in neutrino physics, it is possible that some of these effects may leave an imprint on data. Thus, it is worthwhile to investigate different physics scenarios beyond the SM and study their impact on oscillation probabilities. 
In the present work, we invoke damping effects (including decoherence and decay) on oscillation probabilities and study their implications on the LGI.

\vspace*{.5cm}

\end{titlepage}

\section{Introduction}

A succession of seminal results from low energy experiments involving neutrinos from a variety of sources have established the phenomenon of neutrino flavour oscillations among the three active flavours on a firm footing~\cite{deSalas:2020pgw
}. 
The parameters that govern three flavour oscillations are the three mixing angles ($\theta_{12}, \theta_{23}, \theta_{13}$), one phase ($\delta$) and the two mass-squared splittings ($\Delta m^2_{21}$ and $\Delta m^2_{31}$). While some of the parameters have been measured very 
precisely (such as $\sin^2 \theta_{12}$, $\sin ^2 \theta_{13}$, $\Delta m^2_{21}$, $|\Delta m^2_{31}|$), 
a few of these parameters require better measurements. 
The first of these is the CP phase, $\delta$. While  two long baseline accelerator neutrino experiments, T2K~\cite{T2K:2019bcf} and NOvA~\cite{NOvA:2021nfi} 
have reported measurement of $\delta$, there is some tension in the preferred value of $\delta$. It is not clear if the value of the mixing angle $\theta_{23}$ is maximal or non-maximal (in which case establishing the correct octant is an important question). Finally, the sign of $\Delta m^2_{31}$ which relates to the issue of neutrino mass ordering (normal ordering (NO) if $\Delta m^2_{31} > 0$ and inverted ordering (IO) if $\Delta m^2_{31} < 0$) is unknown. In addition, we should also mention that there are some unresolved anomalies of high statistical significance, notably the event excesses seen in  LSND ($3.8 \sigma$ excess of $\nu_e$ events)~\cite{LSND:2001aii},  MiniBooNE ($4.7 \sigma$ excess of $\nu_e$ events and $2.7 \sigma$ excess of $\bar \nu_e$ events)~\cite{MiniBooNE:2018esg,MiniBooNE:2020pnu},  along with the so-called reactor antineutrino anomaly ($6\%$ flux deficit) ~\cite{Mueller:2011nm} which point towards existence of $eV$ scale sterile neutrinos.   The source of the excess of events seen in MiniBooNE has been further investigated by 
MicroBooNE~\cite{MicroBooNE:2021rmx, MicroBooNE:2021sne,MicroBooNE:2021jwr} which reports no such excess.

Neutrinos are massless in the Standard Model (SM)~\cite{Glashow:1961tr,Weinberg:1967tq,osti_4767615}.
Therefore, the experimental observation of neutrino oscillations,  implying that neutrinos are massive, provides a concrete evidence for the requirement of physics beyond the SM. 
Given the plethora of experimental results  over a wide range of baselines and energies,  
it is evident that neutrinos exhibit coherence over very  vast distance scales.
It is of considerable interest to understand subtle aspects pertaining to the theory of neutrino oscillations and 
coherence properties of neutrinos~\cite{Eliezer:1975ja, 
   Fritzsch:1975rz, 
  Nussinov:1976uw,
     Bilenky:1978nj, 
  Kayser:1981ye,
  Giunti:1991ca, 
  Kiers:1995zj, 
  Kiers:1997pe, 
  Giunti:1997wq, 
  Grimus:1998uh,
Cardall:1999ze,
  Akhmedov:2009rb,
    Kayser:2010pr, 
     Naumov_2010, 
  Akhmedov:2010ms,
 Akhmedov2012, 
 Egorov:2019vqv}.

Current  neutrino oscillation experiments have established neutrino mass and mixing to be the dominant mechanism of flavour conversion of neutrinos, thereby ruling out various other scenarios. However, new physics can give rise to subdominant effects which could be detected at future experiments.  
Such new physics scenarios in conjunction with standard oscillations could even impact the measurement of parameters governing oscillations. In the present work, we are interested in a particular category of such effects which can leave their imprints on the neutrino oscillations in the form of damping terms. Several scenarios can lead to damping of oscillations~\cite{Blennow:2005yk}. Damping effects in neutrino oscillations can originate from 
intrinsic decoherence in the wave packet treatment~\cite{Kiers:1997pe, 
Giunti:1997wq,
Grimus:1998uh,
Cardall:1999ze,Kayser:2010pr,  Naumov_2010, 
  Akhmedov:2010ms,
 Akhmedov2012
},
invisible neutrino decay~\cite{Bahcall:1972my,
Barger:1981vd,
Valle:1983ua,
Barger:1998xk,
Pakvasa:1999ta,
Barger:1999bg,
Lindner:2001fx}, 
neutrino absorption~\cite{DeRujula:1983ya}, 
oscillation into sterile neutrino~\cite{Strumia:2002fw,Maltoni:2004ei} or via 
quantum decoherence~\cite{Gago:2000qc, Ohlsson:2000mj,  Morgan:2004vv,Hooper:2005jp,Anchordoqui:2005gj, 
Farzan:2008eg,
Farzan:2008zv,
Mehta:2011qb,
Kersten:2015kio,
Rasmussen:2017ert,
Coloma:2018idr,
Gomes:2018inp,
Carpio:2018gum,
deGouvea:2020hfl,
Gomes:2020muc,
Nieves:2020jjg,
Stuttard:2020qfv,
Porto-Silva:2021ael}.

While, quantum mechanics as a theory has been immensely successful, 
 questions regarding applicability of quantum mechanics to macroscopic systems have been discussed. 
 In their seminal work, Einstein, Podosky and Rosen~\cite{epr}, questioned, for the first time, if quantum mechanics is the complete description of Nature. 
Later on, Bell~\cite{Bell:1964fg}, in a pathbreaking work, using ideas of local realism, proposed a quantitative way to establish distinction between classical and quantum correlations between spatially separated systems. Since then, various experiments have been proposed to test violation of Bell's inequalities and carried out in various disciplines~\cite{speakable}. 
In order to test the local hidden variable theories, there was another profound development by Leggett and Garg~\cite{Leggett:1985zz} (see \cite{emary2013leggett} for a review) who proposed a set of measurements on a single system at different times. 
These are known as the Leggett-Garg inequalities (LGI). LGI has been studied in the context of non-hermitian  PT symmetric dynamics~\cite{ipsika2021,Karthik:2019wbp}, 
markovian quantum dynamics~\cite{Chanda:2018vmf,Ghosh:2021ncf}.  %
Tests of LGI have been carried out using  photons, electrons, or nuclear spins for which coherence length is limited~\cite{Zhang:2020dva,speakable}. 

A distinguishing feature of neutrino oscillations  is that 
 coherence length   (the length over which interference occurs and oscillations may be observed) extends over very vast length scales.  
 Two of the neutrino oscillation experiments,  MINOS~\cite{Formaggio:2016cuh} and Daya Bay~\cite{Fu:2017hky} 
 have analysed their data   and reported large ($\sim 6 \sigma$) violation of LGI. 
Most of the studies pertaining to violation of LGI in neutrino oscillations have been carried out using standard oscillation framework considering two or three neutrino flavours  but with different dichotomic variables~\cite{Gangopadhyay:2013aha, Gangopadhyay:2017nsn, Naikoo:2017fos,  Richter:2017toa,  Naikoo:2018amb,Naikoo:2019eec}. 
There are also studies using other measures of quantum correlations such as Bell's inequalities or quantum witness~\cite{Alok:2014gya, Dixit:2018kev, Dixit:2018gjc,  Dixit:2019swl, Song:2018bma, Ming:2020nyc, Blasone:2021cau}.
In recent studies, the role of new physics in the context of LGI and neutrino oscillations has been explored~\cite{Shafaq:2020sqo,Sarkar:2020vob}. 
In an earlier work~\cite{Shafaq:2020sqo}, two of the authors of the present work examined the role played by non-standard interactions on the violation of LGI in neutrino oscillations
 and it was shown that new physics could induce large enhancement in violation of LGI parameter. 
In the present article, our aim is to investigate violation of LGI in neutrino oscillations and highlight the role played by those scenarios that lead to damping signatures (including decoherence and decay). 

In order to demonstrate the ideas in the present article, we shall restrict ourselves to two flavour neutrinos and assume that they propagate in vacuum. The mapping of two flavour neutrinos as a two level quantum system is explicitly shown in~\cite{Mehta:2009ea}. This allows us to clearly examine the impact of different damping effects on the violation of LGI. In addition, in many   experimental scenarios, the three flavour formula reduces to an effective two flavour formula~\cite{Bilenky:1978nj} for instance, in the long baseline accelerator experiments, one can set the small parameters ($\theta_{13}$ and $\Delta m^2_{21}$) to zero and obtain effective two flavour formula for muon disappearance probability. The problem  then reduces to $\nu_\mu \to \nu_\tau$ oscillations described by $\theta_{23}$ and $\Delta m^2_{32}$. Likewise, for short baseline reactor experiments, one is not sensitive to oscillations developed by $\Delta m^2_{21}$ and this parameter can be set to zero to obtain effective two flavour formula for the electron antineutrino disappearance probability driven by $\theta_{13}$ and $\Delta m^2_{31}$.

The article is structured as follows. We begin with the standard plane wave description of neutrino oscillations and then go on to describe the damping effects in Sec.~\ref{sec:2}. We describe LGI in neutrino oscillations in Sec.~\ref{sec:3}. We present our results on LGI violation quantified in terms of parameter $K_4$ in Sec.~\ref{sec:4}. Finally, we conclude in Sec.~\ref{sec:5}.

\section{Damped neutrino oscillations}
\label{sec:2}

We will first briefly review the standard treatment of neutrino oscillations   for the undamped case. 
Neutrino oscillations arise from the phenomenon of neutrino mixing. 
In the standard plane-wave theory of neutrino oscillations, it is assumed that the neutrinos (created or detected together with a charged antilepton) 
are described by the flavour neutrino state 
\bea
| {\nu_\alpha} \rangle &=& \sum_{k=1}^{3} U ^\ast_{\alpha k} ~ | \nu_k \rangle 
\eea
where $U $ is the leptonic mixing matrix. 
Since the mass  states  have definite mass ($m_k$) and definite energy ($E_k$), they evolve in time as plane
waves
\bea
| {\nu_k} (t) \rangle &=&  e^{-iE_k t} ~ | \nu_k (t=0) \rangle 
\eea
  The standard theory of neutrino oscillation relies on following assumptions
\begin{itemize}
\item 
The flavour states are the neutrinos produced or detected in charged-current weak interaction processes.
\item The mass states $ | \nu_k \rangle$  have equal momentum.
\item The time can be replaced by distance between source and detector. 
\end{itemize}
Thus, for the undamped case, the probability for three-flavour neutrino oscillations in vacuum can be expressed as
\begin{eqnarray}
P_{\alpha \beta} 
& = & \left| \langle \nu_\beta |
U \, \text{diag} \left( 1 ,  \exp \left( -{\rm i}\frac{\sdm L}{2 E} \right),
 \exp \left( -{\rm i}\frac{\ldm L}{2 E} \right)
\right) \, U^\dagger | \nu_\alpha \rangle \right|^2 \nonumber \\ & = &
\sum\limits_{i,j=1}^{3} {\underbrace{U_{\alpha j} \, U_{\beta j}^* \, U_{\alpha
i}^* \, U_{\beta i}}_{J_{ij} ^{\alpha \beta}}} \, \exp(- {\rm i} 2 \Delta_{ij} )
\end{eqnarray}
\noindent 
where $U$ is the leptonic mixing matrix in vacuum~\cite{mns,Gribov:1968kq}, $J_{ij} ^{\alpha \beta}$ is the Jarlskog factor~\cite{Jarlskog:1985ht},  $\Delta_{ij}
=\Delta m_{ij}^2 L/(4 E)$ denotes the oscillation phase with $\Delta m_{ij}^2
\equiv m_i^2 - m_j^2$ being the mass-squared differences.  After some simplification, we can express the above expression as 
\begin{eqnarray}
P_{\alpha\beta} &=& {\underbrace{
\sum_{i,j = 1}^3 \re{J_{ij}^{\alpha\beta}} -
4 \sum_{1\leq i<j \leq 3} \re{J_{ij}^{\alpha\beta}} \sin^2 (\Delta_{ij}) }_{\textrm{CP conserving}}} -
{\underbrace{
2 \sum_{1\leq i<j \leq 3} \im{J_{ij}^{\alpha\beta}}\sin (2\Delta_{ij}) }_{\textrm{CP violating}}}
\nonumber \\
\label{vacprob}
&=&\sum_{i=1}^3 J_{ii}^{\alpha\beta} +
2 \sum_{1\leq i < j \leq 3}
|J_{ij}^{\alpha\beta}| \cos(2\Delta_{ij}+\arg J_{ij}^{\alpha\beta})
\end{eqnarray} 
Eq.~\ref{vacprob} is used in analyses of the experimental data on neutrino oscillations in vacuum.

Let us now consider damping effects that can arise due to variety of reasons. Some examples of damping signatures which have been widely studied in literature in the context of neutrino oscillations are given in Table~\ref{tab:damping}. It is interesting to note that we can unify the description of different  damping cases using  damping factors of the form~\cite{Blennow:2005yk}
\bea
\label{dfactor}
D_{ij} &= &\exp\left(-\kappa_{ij}\frac{|\Delta m_{ij}^2|^\xi L^\beta}
{E^\gamma}\right)
\eea
where, we assume $D_{ij}=D_{ji}$. In Eq.~\ref{dfactor}, $\kappa_{ij} \ge 0$ is a non-negative damping coefficient matrix, and
$\beta$, $\gamma$, and $\xi$ are numbers that describe the
``signature'', \ie, the $L$ ($\beta$) and $E$ ($\gamma$) dependencies
as well as the dependence on the mass-squared differences. Depending on the value of $\xi$, there can be two cases:  
\begin{description}
\item[${\xi>0}$:] only the oscillatory terms are expected to be damped, since $\Delta m_{ii}^2 = 0$
\item[${\xi=0}$:] the oscillation probability could be damped (depending on $\kappa_{ij}$), since terms  independent of the oscillation phases are affected
\end{description}
In presence of damping, the probability can be expressed as 
\begin{eqnarray}
P_{\alpha\beta} &=& \sum\limits_{i,j=1}^{3}
U_{\alpha j} \, U_{\beta j}^* \, U_{\alpha
i}^* \, U_{\beta i} \, \exp(- {\rm i} 2 \Delta_{ij} ) D_{ij} \nonumber \\
&=&\sum_{i=1}^3 J_{ii}^{\alpha\beta} D_{ii} +
2 \sum_{1\leq i < j \leq 3}
|J_{ij}^{\alpha\beta}| D_{ij}\cos(2\Delta_{ij}+\arg J_{ij}^{\alpha\beta})
\label{damping}
\end{eqnarray}
As $D_{ij} \rightarrow 1$, the above expression reduces to Eq.~\ref{vacprob}.

We now discuss the various physically well-motivated damping scenarios 
in turn below (see Table~\ref{tab:damping}) :

 \begin{enumerate}

\item {\bf{Intrinsic wave packet decoherence}}

This results if we do not assume that mass states propagate as plane waves. 
  In the wave packet treatment of neutrino oscillations, it is inferred that there exists a  distance called ``the coherence length"
    beyond which the interference of different mass states is not observable. 
   This is due to the different group velocities of neutrinos carrying different masses that causes a separation of their wave
packets when they arrive at the detector. 
For a detailed discussion of this scenario, see\cite{Kiers:1997pe, 
Giunti:1997wq,
Grimus:1998uh,
Cardall:1999ze,Kayser:2010pr,  Naumov_2010, 
  Akhmedov:2010ms,
 Akhmedov2012
}.

\begin{table}[htb!]
\begin{center}
\begin{tabular}{cp{3.5cm}ll}
\hline
&&\\
S. No. &\begin{tabular}{l} Damping Scenario \end{tabular} &   $D_{ij} = \exp\left(-\kappa_{ij}\frac{|\Delta m_{ij}^2|^\xi L^\beta}
{E^\gamma}\right)$  & $\kappa$ (units) \\\
&&\\
\hline
& {
\sl{Decoherence like}}  &\\
& $\xi \neq 0$ &\\
1&\begin{tabular}{l} Intrinsic wave \\packet decoherence \end{tabular} &
$\exp \left( - \sigma_E^2 \dfrac{(\Delta m_{ij}^2)^2 L^2}{8E^4} \right)$ &
$\dfrac{\sigma_E^2}{8}$ ($\mathrm{GeV ^{2}}$)    \\
2 &\begin{tabular}{l} Quantum \\ decoherence \end{tabular} &
 $\exp \left( - \kappa \dfrac{(\Delta m_{ij}^2)^2 L^2}{E^2} \right)$ &
 $\kappa$ (dimensionless)   \\
 \hline 
 & {\sl{Decay like }} &\\
 & $\xi = 0$ &\\
  3 &\begin{tabular}{l} Invisible neutrino \\ decay \end{tabular} & $\exp \left( - \kappa \dfrac{L}{E} \right)$ &
$\kappa$ ($\mathrm{GeV \cdot km^{-1}}$)  \\
 4 &\begin{tabular}{l} Oscillations into \\ sterile neutrino  \end{tabular}& $ \exp \left( - \epsilon \dfrac{L^2}{(2E)^2} \right)$ &
$\epsilon$ ($\mathrm{eV ^{4}}$)   \\
 5 &\begin{tabular}{l}  Neutrino \\absorption \end{tabular} & $\exp \left( - \kappa L E \right)$ &
$\kappa$   ($\mathrm{GeV^{-1} \cdot km^{-1}}$)   \\
\hline
\end{tabular}
\end{center}
\caption{\label{tab:damping} Damping scenarios considered in the present work.}
\end{table}

For this damping signature, using  averaging over Gaussian wave packets%
, we have 
%
\bea
D_{ij}  &=& \exp \left[ - \left( \dfrac{L}{L_{ij}^{\mathrm{coh}}}
\right)^2 \right] =
 \exp \left(  - \dfrac{\sigma_E^2}{8}  \dfrac{(\Delta m_{ij}^2)^2 L^2}{ E^4} \right)
\label{intrin}
\eea
which takes into account the coherence of the contributions of different mass
eigenstate wave packets. 
 Here $L_{ij}^{\mathrm{coh}} = 4 \sqrt{2} \sigma_x
E^2/|\Delta m_{ij}^2|$ and $\sigma_x$ is the spatial coherence width of the mass eigenstate wave packet. 
$\sigma _x$ depends on the spatial as well as 
 temporal coherence widths of both the production and detection processes. 
 $\sigma_E = 1/(2
\sigma_x)$ is the wave packet spread in energy.  The parameter $\sigma_E$ could be different for different experiments.
In most oscillation experiments, the condition $\sigma_x <<    {4 \pi E}/{\Delta m^2_{ij}} \equiv  L_{ij}^{\mathrm{osc}} $ is  satisfied. 
{{By comparing Eqs.~\ref{dfactor} and
\ref{intrin}, we note that $\kappa_{ij} = \sigma_E^2/8$, $\beta=2$, $\gamma=4$, and $\xi=2$. 
}}

If we consider the case corresponding to  $\xi > 0$. 
For this case, the diagonal terms, $D_{ii} = 1$. 
The probability for two flavour case can be expressed as 
 \bea
P_{\alpha \beta} &=& \delta_{\alpha \beta} + \dfrac{1}{2} \left(1 - 2\delta_{\alpha \beta}\right) \sin^2(2\theta)\left[
1 - D_{21} \cos(2\Delta_{21})\right]
\label{eq:1}
\eea
Note the dependence of probability on the sign of $\cos (2 \Delta_{21})$ term. {{This explains the fact that near the location of the peak, we get suppression and near the location of the dip, we get enhancement. }} This case is  {\sl{``decoherence-like" (probability conserving)}}. 
In the limiting case, $D_{21} \to 0$, we obtain the expression 
 \bea
P_{\alpha \beta} &\to& \delta_{\alpha \beta} \left[1 - \sin^2 (2 \theta)\right] + \dfrac{1}{2} \sin^2(2\theta)
\eea
which corresponds to the case of averaged oscillations.  

 The effect of neutrino wave packet decoherence on the two flavour oscillation probability is shown in Fig.~\ref{fig_prob_all}.  
 From the figure, it is clear that wave packet decoherence leads to  damping of the oscillatory term only.
 It should be noted that this particular decoherence signature is  related to processes affecting production and detection.

\begin{figure}[t!]
\centering
\includegraphics[width=0.7\linewidth]{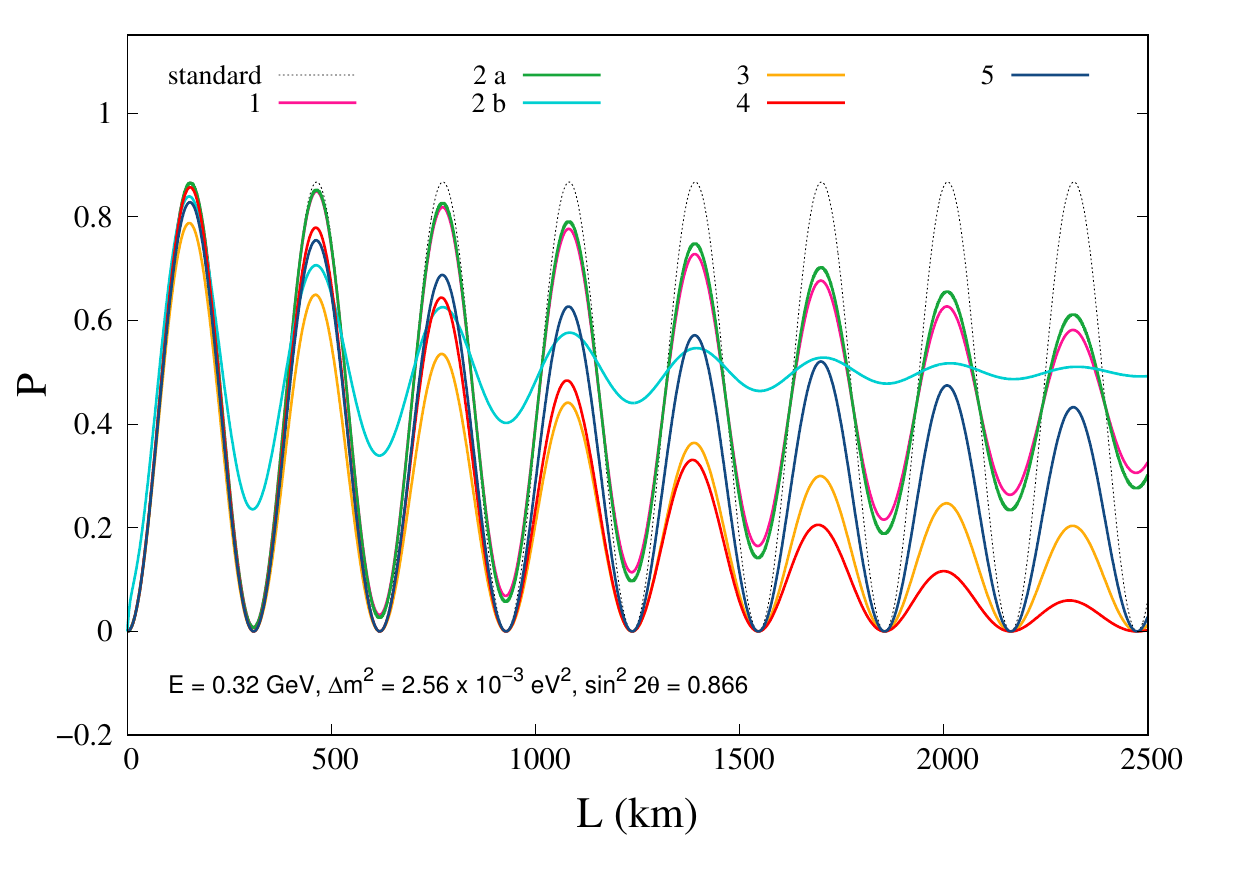}
\caption{\footnotesize{The two flavour oscillation probability corresponding to various damping cases
 and comparison with the standard case.
 }}
\label{fig_prob_all}
\end{figure}

\item {\bf{Quantum decoherence}}

If neutrino system is coupled to an environment, one may encounter effects due to quantum decoherence. 
Quantum decoherence effects in neutrino propagation are introduced using the Liouville-Lindblad formalism for open quantum systems~\cite{Lindblad:1975ef,
Ellis1984381,Banks:1983by,Liu:1993ji,Benatti:2000ph,Mavromatos:2007hv}.  The density matrix $\rho$ describing the neutrino flavour evolves according to 
\begin{equation}
\label{EqLiouvilleLindbladDef}
 \frac{\partial\rho}{\partial t} = -i \left[\ham,\,\rho\right] + \mathcal{D}\left[\rho\right] ~
\end{equation}
where the Hamiltonian, $\ham$, is responsible for the usual unitary evolution, and the extra term, $\mathcal{D}\left[\rho\right]$, for non-unitary evolution, \ie, decoherence. 
We describe all  possible decoherence cases in the context of two flavour neutrino oscillations in Appendix A.

In what follows, we consider two cases :

\begin{itemize}

\item[{\bf{(a): }}] In Ref.~\cite{Ohlsson:2000mj}, it was shown that gaussian uncertainties in the neutrino energy and pathlength, when averaged over, can lead to modifications of the standard neutrino oscillation probability which are similar in form to those coming from this scenario of quantum decoherence. This is the simplest of the decoherence cases listed in  Appendix A (see case 2 which corresponds to the minimal decoherence scenario). 
In this particular case~\cite{Ohlsson:2000mj}, the oscillation probability becomes
\bea
P_{\alpha \beta} &=&
  \frac{1}{2}\sin^2 (2 \theta) \left[1 -   \exp \left(  - \kappa  \frac{(\Delta m^2_{21})^2 L^2}{E^2} \right) \cos \left(\frac{\Delta m^2_{21} L}{2 E}\right)\right]
  \label{eq:5}
\eea
Note that Eq.~\ref{eq:5} is similar in form to Eq.~\ref{eq:1} which is in accord with the main result of~\cite{Ohlsson:2000mj}.

\item[{\bf{(b): }}]    
Let us look at another case (see case 5 in Appendix A) for which $d = \beta = 0$ and which violates energy conservation. 
The two flavour oscillation probability in this case is given by

\begin{eqnarray}
    P_{\alpha \beta}&=&\dfrac{1}{2}\left[1-e^{-2\delta L} \cos^2{2\theta}+e^{-(a+\alpha) L}\sin^2{2\theta}\left\{ -\cos (\Omega L)+\frac{(a-\alpha)}{\Omega}\sin (\Omega L) \right\} \right] \nonumber\\
\end{eqnarray}
where $\Omega = \sqrt{4(\omega^2 - b^2) - (a - \alpha)^2} $ and $\omega = \Delta m^2_{21}/ 4E$.

\end{itemize}

The two flavour oscillation probability for the two cases of quantum decoherence mentioned above is depicted in Fig.~\ref{fig_prob_all} as cases 2(a) and 2(b). It can be noted that the case 2(a) follows similar behaviour as case 1.  
The case 2(b) is quite distinct from 2(a). 
{{We can note that as $ L \to \infty$, probability approaches the steady state value of $1/2$ only in case 2(b) and not in case 2(a). }}

\item {\bf{Invisible neutrino decay}}

This corresponds to decay of neutrinos into particles invisible to the detector  (see, \eg,
\cite{Bahcall:1972my,Barger:1981vd,Valle:1983ua,Barger:1998xk,Pakvasa:1999ta,Barger:1999bg}). As a result, the  three flavour unitarity is lost. In this
case, the neutrino evolution is described by an effective Hamiltonian
\begin{equation}
\ham_{\rm eff} = \ham - {\rm i}\Gamma
\end{equation}
where $\Gamma = \diag(a_1,a_2,a_3)/2$ in the neutrino mass eigenstate
basis, $a_i = \Gamma_i/\gamma_i$, $\Gamma_i$ is the inverse
life-time of a neutrino of the $i^{th}$ mass eigenstate in its own rest frame,
and $\gamma_i  = E/m_i$ is the time dilation factor. We note that
$\ham$ and $\Gamma$ are both diagonal in the neutrino mass eigenstate
basis. 

In this case, the  damping factor in the oscillation probability can be written as 
\begin{equation}
\label{equ:decay}
D_{ij} = \exp\left(-\frac{\kappa_i + \kappa_j}{2} \dfrac{L}{E}\right) 
\end{equation}
where $\kappa_i = \Gamma_i m_i$.

{{Thus, for neutrino decay, the we have $\kappa_{ij} = (\kappa_i +
\kappa_j)/2$, $\beta = \gamma = 1$ and $\xi = 0$. }} 
The damping factor $D_{ij}$ for {\sl{``decay-like scenario"}} can be written as  
\begin{equation}
\label{pbviol}
D_{ij} = A_i A_j 
\end{equation}
where $A_i = \exp(-\kappa_i L^\beta/E^\gamma)$ is only dependent on
the $i^{th}$ mass eigenstate. 
For the two flavour scenario, we can write
\begin{eqnarray}
P_{\alpha\alpha}
&=&
A^2\left[(c^2+\zeta s^2)^2 - \zeta \sin^2(2\theta)
\sin^2\left(\frac{\Delta m_{21}^2 L}{4 E}\right)\right] \\
P_{\alpha\beta}
&=&
\frac{1}{4} A^2 \sin^2(2\theta)\left[1+\zeta ^2 - 2\zeta
\cos\left(2\frac{\Delta m_{21}^2 L}{4 E}\right)\right]
\label{eq:2}
\end{eqnarray}
where $A = A_1$ and $\zeta = A_2/A_1$. 
The total probability is 
\begin{equation}
\label{dprobtot}
P_{\alpha\alpha} + P_{\alpha\beta} =
A^2\left[c^4 + \zeta^2 s^4 +
\frac{1}{4}\sin^2(2\theta)(1+\zeta)^2\right] \leq 1
\end{equation}
which is not conserved in general.  This is depicted in Fig.~\ref{fig_prob_all} and \ref{fig_prob2}.  The loss of unitarity can be noted from Fig.~\ref{fig_prob2}.

\begin{figure}[htb!]
\centering
\includegraphics[width=0.7\linewidth]{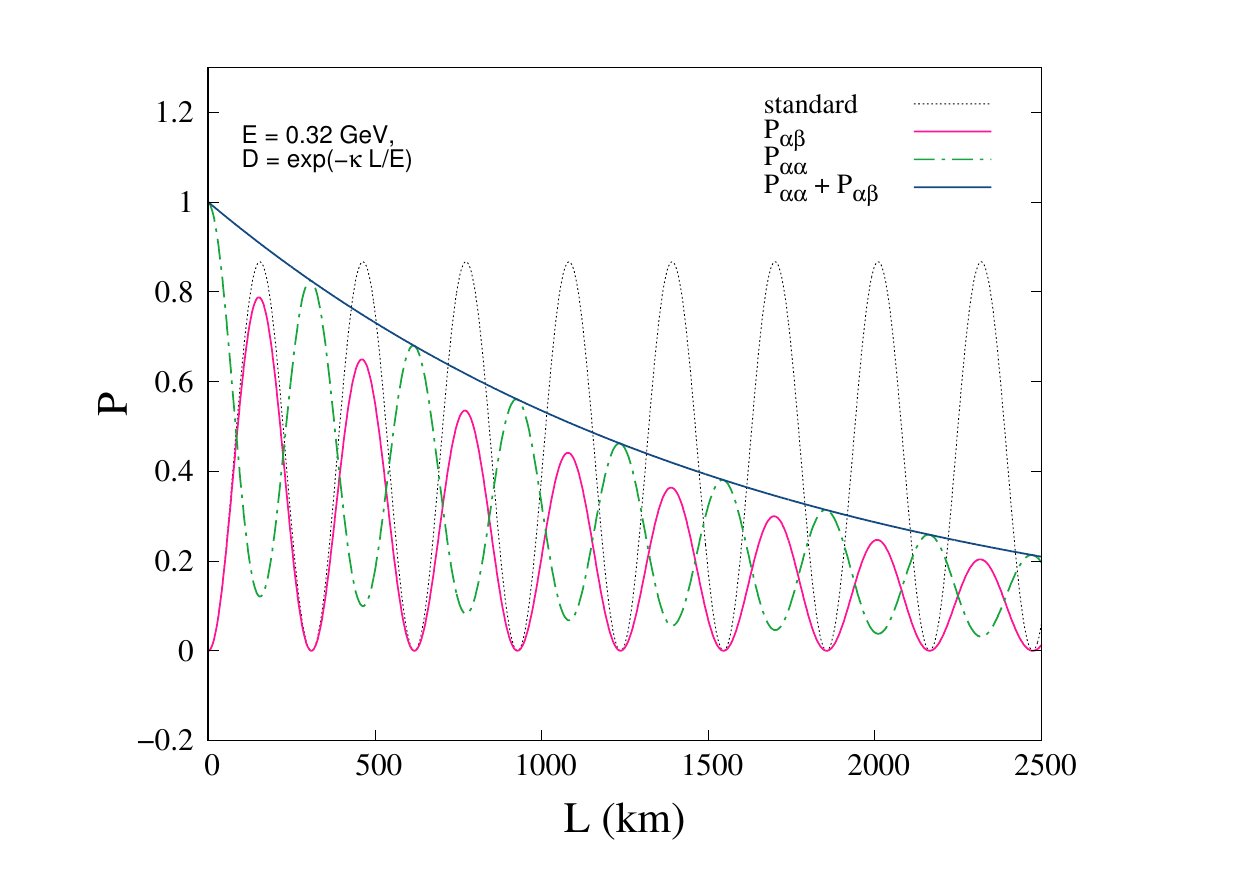}
\caption{\footnotesize{Two flavour oscillation probability corresponding to invisible neutrino decay scenario and comparison with the standard case.
}}
\label{fig_prob2}
\end{figure}

For the simpler  case when $\zeta = 1$, \ie, both mass eigenstates are affected in the same way, 
the above expression 
 reduces to the undamped standard two flavour neutrino oscillation probabilities suppressed by an overall factor of $A^2$ 
 \bea
P_{\alpha \beta} &=& A^2 \left[ \delta_{\alpha \beta} + (1 - 2\delta_{\alpha \beta}) \sin^2(2\theta) \left[1 - \cos(2\Delta)\right] \right]
\label{eq:3}
\eea

\item {\bf{Oscillation into sterile neutrino}}

If there exists one more state in addition to the two active ones, then  
we have terms of the kind, $\sin^2 \Delta_{4i}$ where as before, $\Delta_{ij} =\Delta m_{ij}^2 L/(4E)$.
This results in a loss of unitarity  ($\epsilon$ represents the magnitude of the mixing) coded in the following way,
\bea
 1 - \epsilon \, \sin^2 \left(\frac{\Delta m_{4i}^2 L}{4E} \right)
\simeq 1 - \epsilon \left( \frac{\Delta m_{4i}^2 L}{4 E} \right)^2
\simeq \exp \left[
-  \epsilon \left( \frac{\Delta m_{4i}^2 L}{4 E} \right)^2 \right] 
\eea 
In vacuum, we use  the following damping signature~\cite{Strumia:2002fw,Maltoni:2004ei}
\bea
D_{ij} &=& \exp \left( - \kappa_{ij} \frac{L^2}{(2E)^2} \right) = \exp \left( - \dfrac{\epsilon}{4} \frac{L^2}{E^2} \right)  
\label{oscillations}
\eea
where $\epsilon$ contains the information on mixing and $\Delta m^2$
and will be given in units of $\mathrm{eV}^4$. Here, 
the damping coefficient $\alpha$ depends on 
the new mixing angle and the new mass-squared differences, $\Delta
m_{4i}^2$. 
Thus,  comparing Eq.~\ref{oscillations}
with Eq.~\ref{dfactor}, we find that $\kappa_{ij} = \epsilon/4$, $\beta=\gamma=2$,
and $\xi=0$.

The oscillation probability for the two flavour case is 
\begin{equation}
P_{
\alpha \beta} = \exp\left(-\frac{\epsilon L^2}{4 E^2}\right) \left[
 \sin^2 2 \theta \sin^2 \left(\frac{\Delta m_{21}^2 L}{4 E}\right) \right]
 \label{eq:3}
\end{equation}
The two flavour oscillation probability for this case depicted in Fig.~\ref{fig_prob_all}. The loss of unitarity can be noted from Fig.~\ref{fig_prob4}.
Note that the fall off of the sum of probabilities has a distinct behaviour with respect to case 3 (Fig.~\ref{fig_prob2}). 

\begin{figure}[htb!]
\centering
\includegraphics[width=0.7\linewidth]{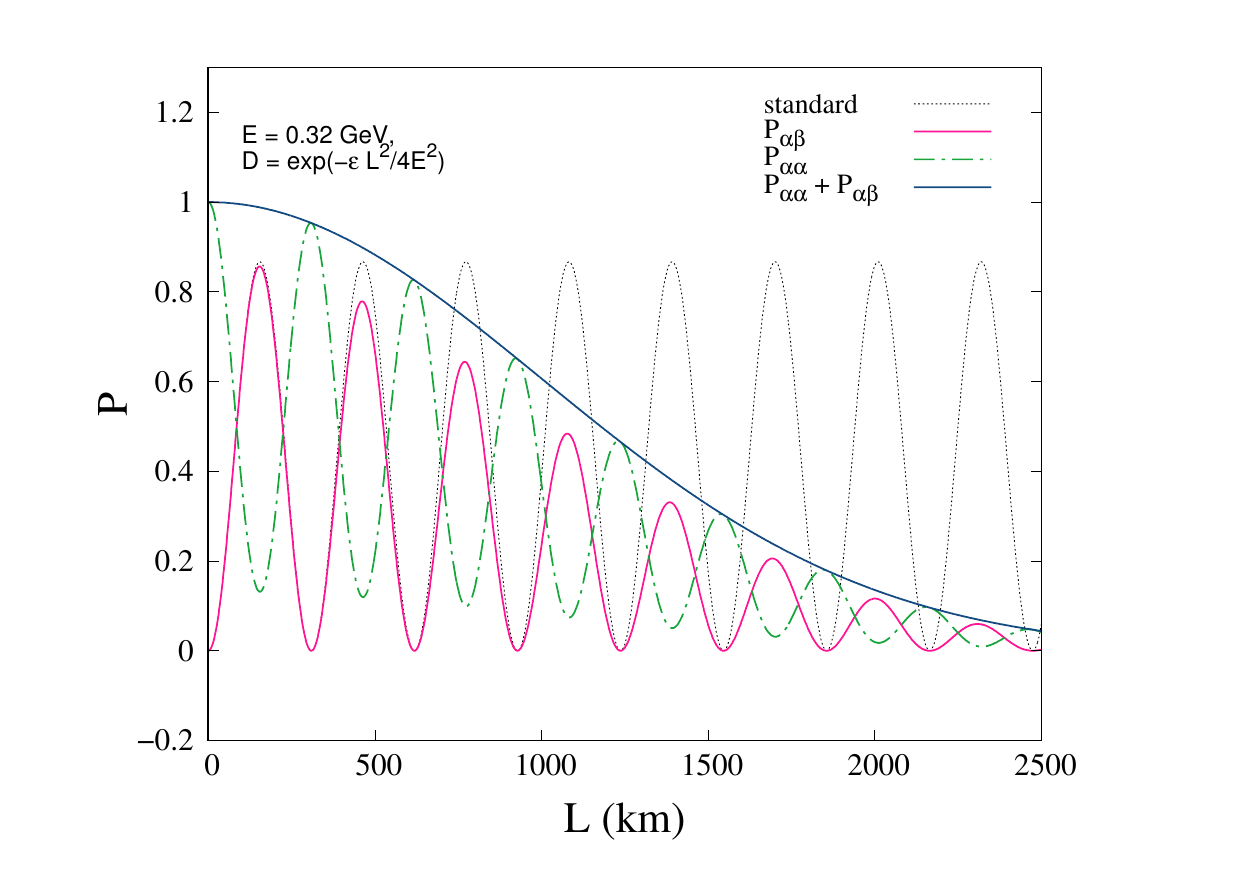}
\caption{\footnotesize{Two flavour oscillation probability corresponding to oscillation into sterile neutrino and comparison with the standard case.} }
\label{fig_prob4}
\end{figure}

\item {\bf{Neutrino Absorption}}

\begin{figure}[htb!]
\centering
\includegraphics[width=0.7\linewidth]{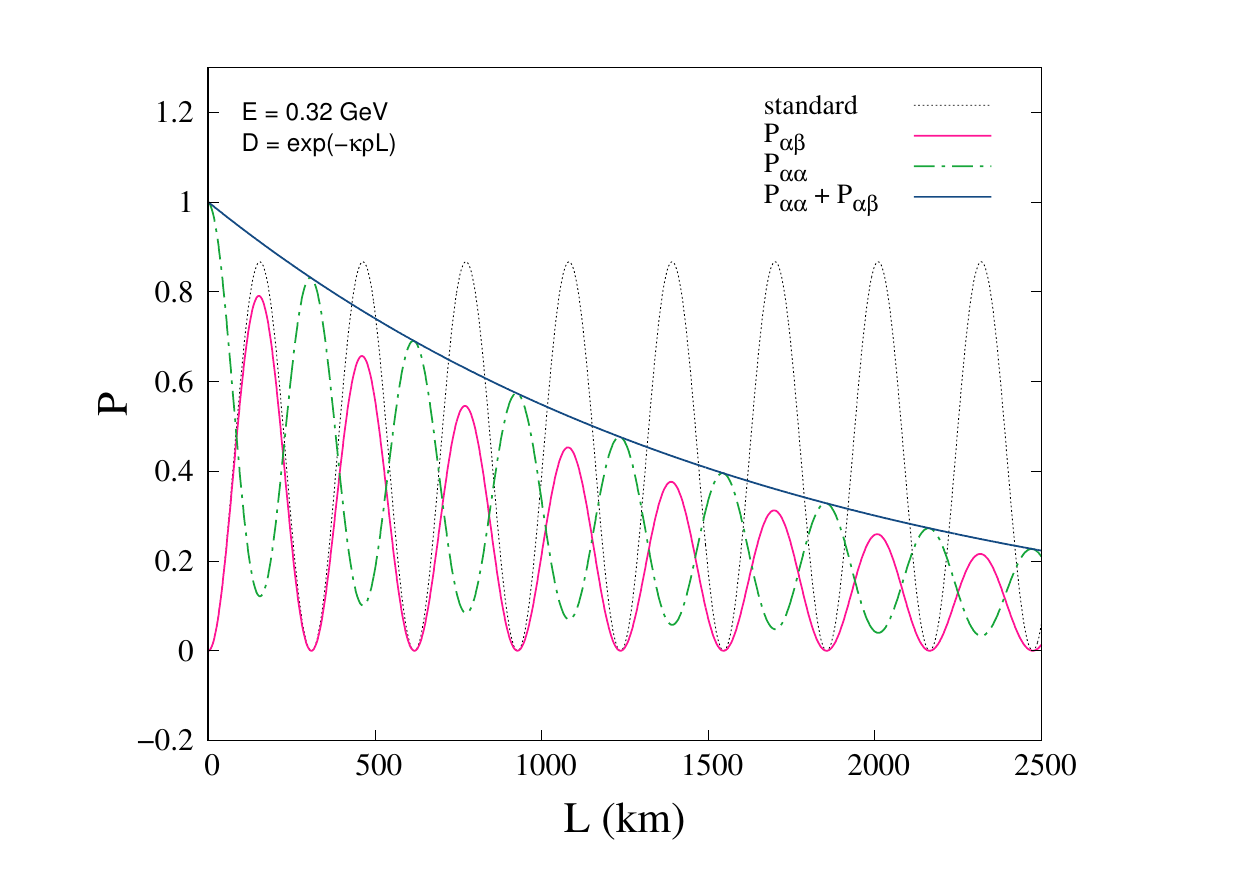}
\caption{\footnotesize{Two flavour oscillation probability corresponding to absorption scenario and comparison with the standard case.
}}
\label{fig_prob5}
\end{figure}  

 Neutrino absorption{{~\cite{DeRujula:1983ya}}} can be described in a similar fashion as neutrino decay. In this case,  the effective Hamiltonian is given by
\bea
\ham_{\rm eff} &=
& \ham - {\rm i}\Gamma
\eea
where $\ham$ is the standard neutrino Hamiltonian in matter, $\Gamma$ is
given by
\begin{equation}
\Gamma = \rho \diag(\sigma_e, \sigma_\mu, \sigma_\tau)/2
\end{equation}
in the flavour eigenstate basis, $\rho$ is the matter density, and
$\sigma_\alpha$ is the absorption cross-section for a neutrino of
flavour $\alpha$. Under the assumption that the cross-sections  are 
small, the eigenstates of $\ham_{\rm eff}$  do not differ much from the orthogonal eigenstates of $\ham$. Thus, the first
order corrections to the eigenvalues of the effective Hamiltonian can be written as
\begin{equation}
\delta E_i^{(1)} = -{\rm i} \Gamma_{ii} =
-{\rm i} \frac \rho 2 \sum_\alpha |U_{\alpha i}|^2 \sigma_\alpha
=
-{\rm i}\frac \rho 2 \sigma_i
\end{equation}
where $\sigma_i$ is an effective cross-section for a neutrino of mass
eigenstate $i$. The neutrino oscillation probability (taking $\rho$ to be constant) is now given by
an expression of the form of Eq.~\ref{damping} with
\begin{equation}
D_{ij} = \exp\left(
-\frac{\sigma_i + \sigma_j}{2} \rho L
\right)= \exp\left(
-\frac{\sigma_i(E) + \sigma_j(E)}{2} \rho L
\right) \, 
\end{equation}  
{{Here $\beta = 1$ and $\gamma$ is equal to
minus the power of the energy dependence of the cross-sections.}}
The two flavour oscillation probability is given by
\begin{eqnarray}
P_{\alpha \beta} &=& \exp\left(-\kappa \rho L\right) \sin^2 2 \theta \sin^2 \left(\frac{\Delta m_{21}^2 L}{4 E}\right)\nonumber\\
P_{\alpha \alpha} &=& \exp\left(- \kappa \rho L\right)\left[1- \sin^2 2 \theta \sin^2 \left(\frac{\Delta m_{21}^2 L}{4 E}\right)\right]
\label{eq:4}
\end{eqnarray}
 The two flavour oscillation probability for this case 
depicted in Fig.~\ref{fig_prob_all}. The loss of unitarity can be noted from Fig.~\ref{fig_prob5}. Note that the fall off of the sum of probabilities has a 
similar behaviour with respect to  case 3 (Fig.~\ref{fig_prob2}) but 
distinct  with respect to  case 4 (Fig.~\ref{fig_prob4}). 
  
  \end{enumerate}

\section{LGI in two flavour neutrino oscillations : recap of the undamped case}\label{sec:3}

Leggett and Garg~\cite{Leggett:1985zz} derived a set of inequalities which provide a way to test the 
 quantumness of a system. Our intuition about the view of macroscopic world relies on 
(see \cite{emary2013leggett} for a review)
 \begin{itemize}
\item[(a)]
 Macroscopic realism (MR) 
 \item [(b)] Non-invasive measurability (NIM) 
 \end{itemize}
 MR  implies that a system remains in a fixed pre-determined state at all times. 
NIM  implies that we can measure this value without disturbing the system.  
Both these assumptions are respected by the classical world. 
However, these are violated in quantum mechanics which is based on superposition principle and  collapse of wave function under measurement. 

\begin{figure}[ht!]
\centering
\includegraphics[width=0.75\linewidth]{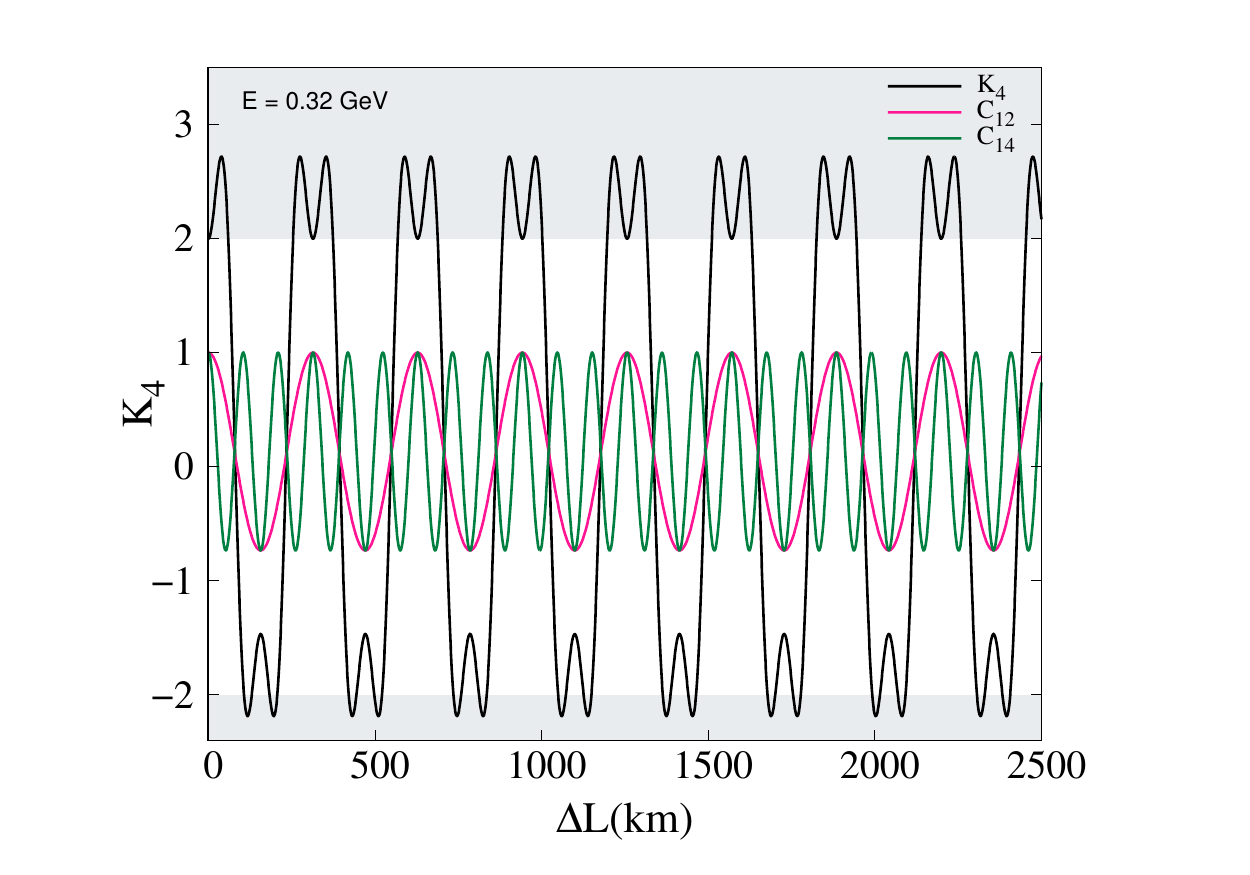}
\caption{\footnotesize{The LGI parameter $K_4$ is plotted for undamped two flavour case.}}
\label{k4_undamped}
\end{figure}

In order to write down an expression for LGI, we require correlation functions $C_{ij} = \langle \hat Q (t_i) \hat Q (t_j) \rangle$ of a dichotomic observable, $\hat Q (t)$ (with realizations $\pm 1$) at distinct measurement times $t_i$ and $t_j$. Here, $\langle \ldots \rangle$ implies averaging over many trials. 
 The two-time correlation function is given by
\bea
C_{ij} &=& \sum _{\hat Q_i \hat Q_j = \pm1} \hat Q_i \hat Q_j \mathbb{P}_{\hat Q_i \hat Q_j} (t_i,t_j)
\eea
{where  $\mathbb{P}_{\hat Q_i \hat Q_j} (t_i,t_j)$ is the joint probability of obtaining the results $\hat Q_i$ and $\hat Q_j$ from successive measurements at times $t_i$ and $t_j$ respectively. }
We can define $K_4$ as 
\begin{eqnarray}
K_4 = C_{12} + C_{23} + C_{34} - C_{14}
\end{eqnarray} 
$K_4$ satisfies the following inequality~\cite{Leggett:1985zz},
\begin{eqnarray}
-2 \leq K_{4} \leq 2 \quad \quad  
\label{genlgi}
\end{eqnarray}
In order to explore LGI in the neutrino sector, we consider a muon neutrino $|\nu_\mu \rangle$ on which measurements are made at times $t_{i}$. The dichotomic observable $\hat Q$ assumes value $+1$ when the system to be found in the muon neutrino flavour state $|\nu_{\mu}\rangle$ and $-1$ if the system is found in the electron neutrino  state $|\nu _{e}\rangle$.  Now, given an initial muon neutrino beam, after time $t$ (or distance $L$), the probability of obtaining $\nu_{e}$
 is given by
\begin{eqnarray}
P_{{\mu} {e}}  &=& \sin^{2} 2\theta \sin^{2}\frac{\Delta m^2 L}{4 E} 
\end{eqnarray}
For the two flavour case, which mimics a two-level quantum system~\cite{Mehta:2009ea}, the $C_{12}$ can be written as~\cite{Gangopadhyay:2013aha}
\begin{eqnarray}
C_{12} &=& \mathbb{P}_{\nu_{e} \nu_{e}}(t_1,t_2)-\mathbb{P}_{\nu_{e} \nu_{\mu}}(t_1,t_2)-\mathbb{P}_{\nu_{\mu} \nu_{e}}(t_1,t_2)+\mathbb{P}_{\nu_{\mu} \nu_{\mu}}(t_1,t_2)   \nonumber
\end{eqnarray}
 {where {{{{$\mathbb{P}_{\nu_{\alpha} \nu_{\beta}} (t_1,t_2)  
 = P_{\mu \alpha} (t_1) P_{\alpha \beta} (t_2)$}}}}  is the 
joint probability of obtaining  neutrino in state $|\nu_\alpha\rangle$ at time 
$t_{1}$ and in state $|\nu_\beta \rangle $ at time $t_{2}$.  
In the ultra-relativistic limit, this time difference translates to the spatial difference $\Delta L = (L_{i} - L_{j})$, where $L_{i}$ and $L_{j}$ are the fixed distances from the neutrino source where the measurements occur. Therefore we have,
\begin{eqnarray}
C_{12} = 1- 2\sin^2 2\theta \sin^2 \left(\dfrac{\Delta m^2 \Delta L}{4 E} \right)
\end{eqnarray}

 \begin{figure}[ht!]
\centering
\includegraphics[width=0.65\linewidth]{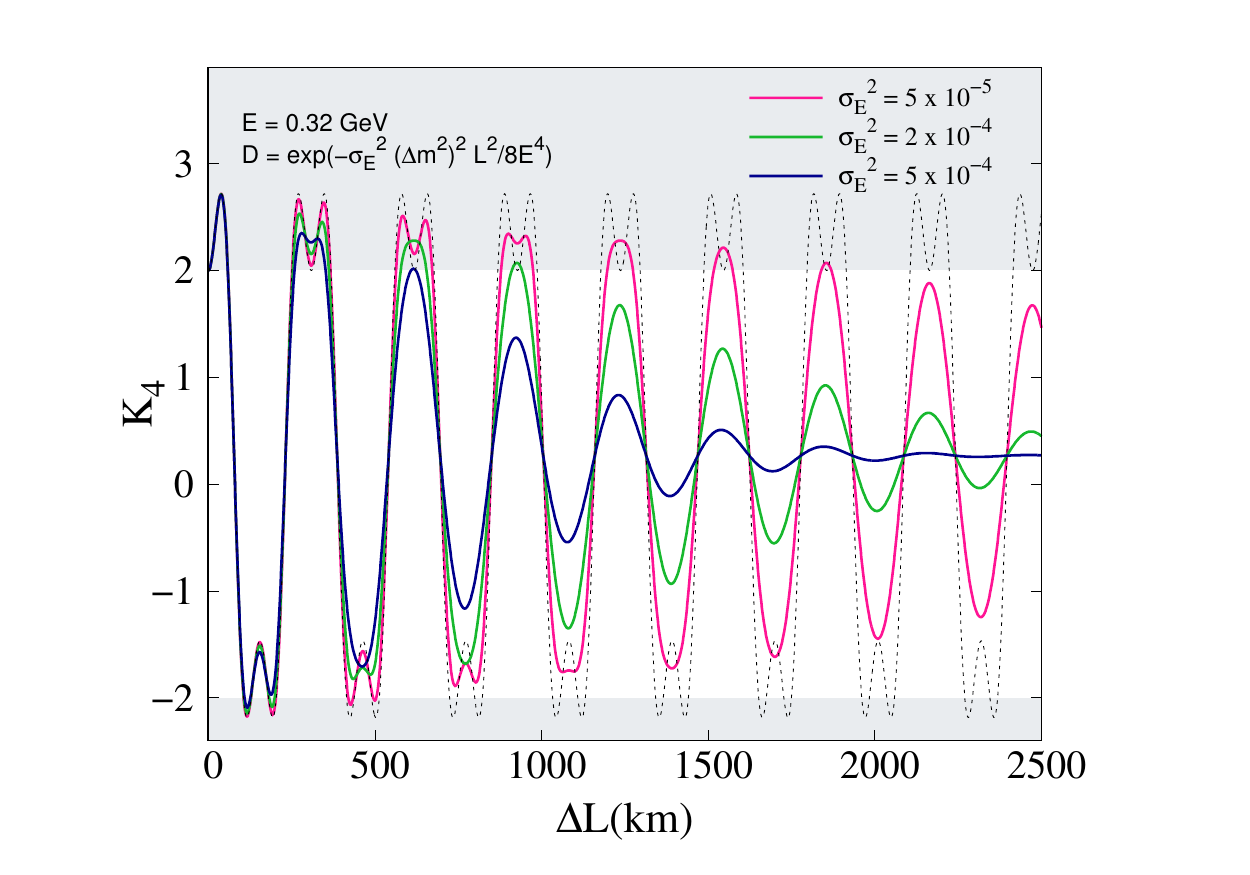}
\caption{\footnotesize{$K_4$ as a function of $\Delta L$ for  wave packet decoherence (case 1).
}}
\label{k4_intrinsic}
\end{figure}  
 
The other correlation functions required to evaluate $K_{4}$, \ie, 
$C_{23} , C_{34}$ and $C_{14}$ can be derived in the same way as $C_{12}$.
Taking $L_4-L_3 = L_3-L_2 = L_2-L_1 = \Delta L$, $K_4$ is given by
\begin{eqnarray}
K_4 &=& 2-2\sin^{2} 2\theta \Big\{3\sin^{2}\frac{\Delta m^2 \Delta L}{4 E}-\sin^{2}3\frac{\Delta m^2 \Delta L}{4 E}\Big\}
\label{k4-co}
\end{eqnarray}
In Fig.~\ref{k4_undamped}, we show the variation of $K_4$ and $C_{ij}$'s as a function of $\Delta L$ for the standard
two flavour case. It can be noted that three of the $C_{ij}$'s are the same, \ie, $C_{12} = C_{23} = C_{34}$,  while $C_{14}$ is different. The interplay between the four terms is responsible for the overall behaviour of $K_4$. The grey shaded region implies quantum regime as it is outside the range defined by $-2 \leq K_4 \leq 2$. Thus, the standard two flavour case adheres to quantum mechanics. 
 Following the procedure laid down for the undamped case, here we evaluate the $C_{ij}$'s
 and $K_4$ for the damping cases discussed below.

\section{Results - Damped oscillations and LGI}
\label{sec:4}

 We consider different scenarios of damped two flavour neutrino oscillations as described earlier. 
We first discuss the case of {\bf{intrinsic wave packet decoherence}}. 
 Using Eq.~\ref{eq:1}, we obtain  $C_{12}$ as 
\bea
C_{12} &=& 1 + \sin^2 (2 \theta) \left[\exp \left(  - \sigma_E^2  \frac{(\Delta m^2)^2 (L_2 - L_1)^2}{8 E^4} \right) \cos \left(\frac{\Delta m^2 (L_2-L_1)}{2 E}\right)-1 \right] \nonumber\\
\eea
This allows us to write down the LGI parameter $K_4$ as
\begin{eqnarray}
K_4 &=& 3 \left[1 + \sin^2 (2 \theta) [\exp \left(  - \sigma_E^2  \frac{(\Delta m^2)^2 \Delta L^2}{8 E^4} \right) \cos \left(\frac{\Delta m^2 \Delta L)}{2 E}\right)-1]\right] \nonumber\\
&& \,- \left[1 + \sin^2 (2 \theta) [\exp \left(  - \sigma_E^2  \frac{9(\Delta m^2)^2 \Delta L^2}{8 E^4} \right) \cos \left(\frac{3\Delta m^2 \Delta L}{2 E}\right)-1] \right]
\end{eqnarray}
$K_4$ is plotted as a function of $\Delta L$ for this case in Fig.~\ref{k4_intrinsic}. The different curves 
correspond to different values of the parameter $\sigma_E$. It can be noted from the plot that 
{{ $K_4 \to 0.268$}}
 as we  go to large values of $\Delta L$.

The next case concerns the scenario of {\bf{quantum decoherence}}. Here we have explored two possibilities. 
We first examine one of the  most  well-studied case of quantum decoherence~\cite{Ohlsson:2000mj}.
Using Eq.~\ref{eq:5}, the correlation function $C_{12}$ can be expressed as
\bea
C_{12} &=& 1 + \sin^2 (2 \theta) \left[\exp \left(  - \kappa  \frac{(\Delta m^2)^2 (L_2 - L_1)^2}{E^2} \right) \cos \left(\frac{\Delta m^2 (L_2-L_1)}{2 E}\right)-1\right]
\eea
The LGI parameter, $K_4$ is given by
\begin{eqnarray}
K_4 &=& 3 \left[1 + \sin^2 (2 \theta) [\exp \left(  - \kappa  \frac{(\Delta m^2)^2 \Delta L^2}{E^2} \right) \cos \left(\frac{\Delta m^2 \Delta L)}{2 E}\right)-1]\right] \nonumber\\
&-& \left[1 + \sin^2 (2 \theta) [\exp \left(  - \kappa  \frac{9(\Delta m^2)^2 \Delta L^2}{E^2} \right) \cos \left(\frac{3\Delta m^2 \Delta L}{2 E}\right)-1] \right]
\end{eqnarray}
\begin{figure}
\centering
\includegraphics[width=0.66\linewidth]{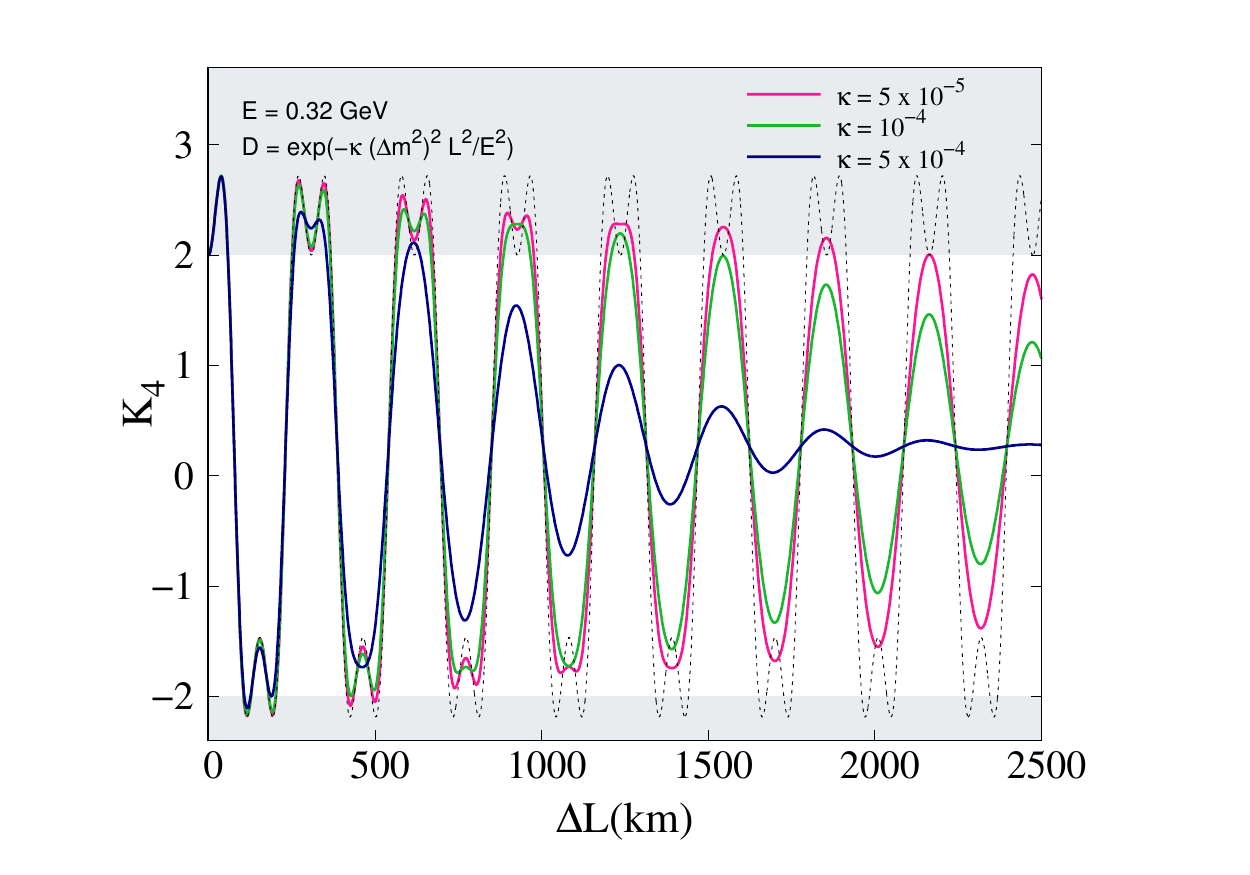}
\caption{\footnotesize{$K_4$ as a function of $\Delta L$ for  quantum decoherence (case 2 (a)). 
This case is similar to case 1.
}}
\label{k4_decoh}
\end{figure}
\begin{figure}
\centering
\includegraphics[width=0.48\linewidth]{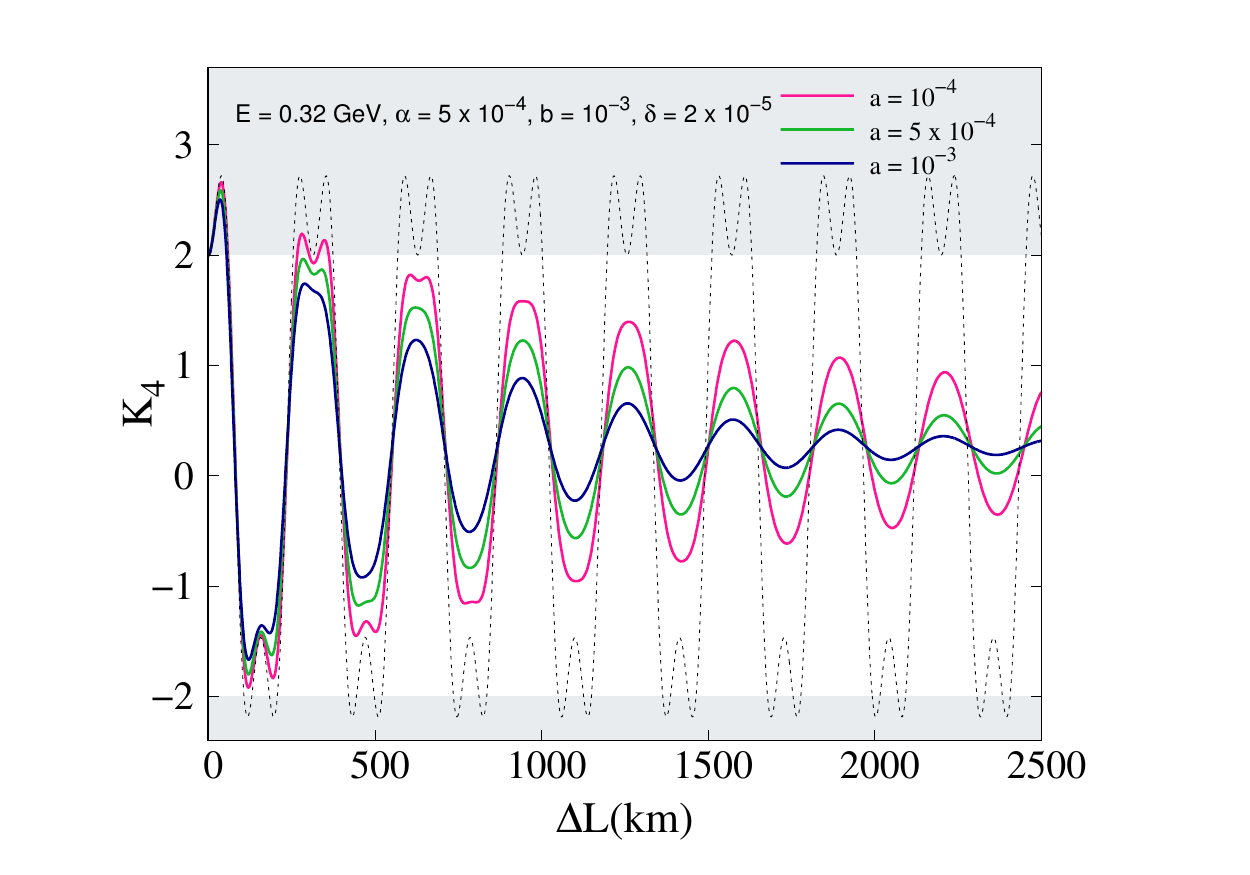}
\includegraphics[width=0.48\linewidth]{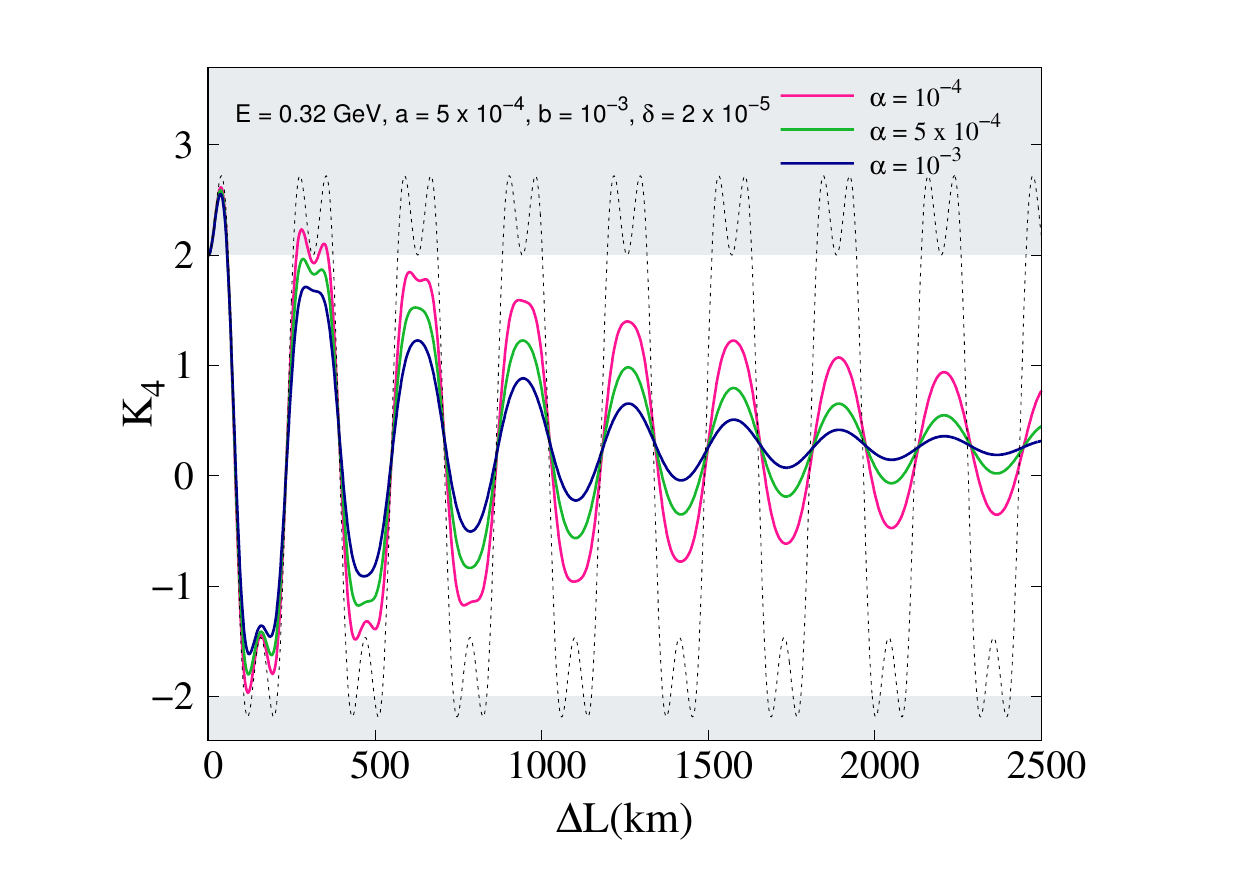}
\includegraphics[width=0.48\linewidth]{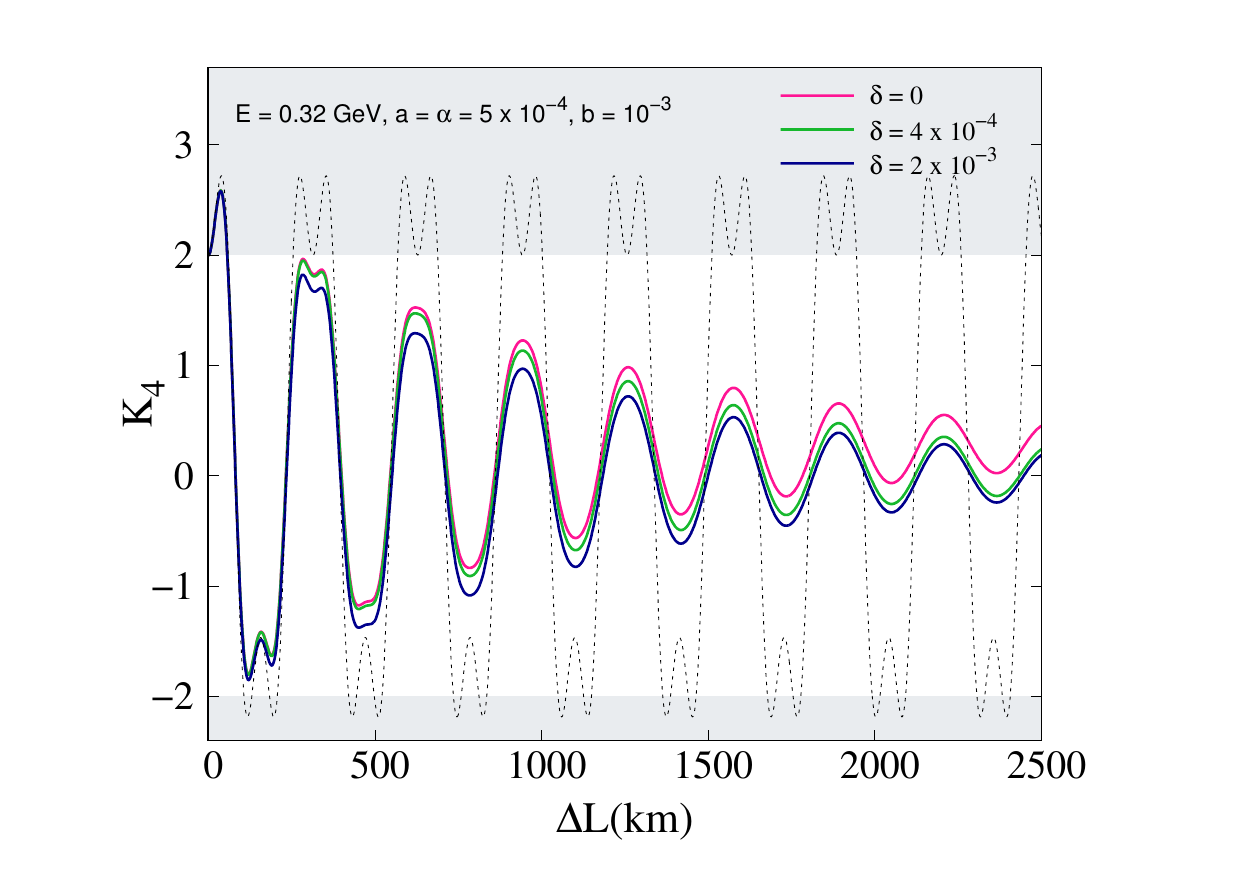}
\includegraphics[width=0.48\linewidth]{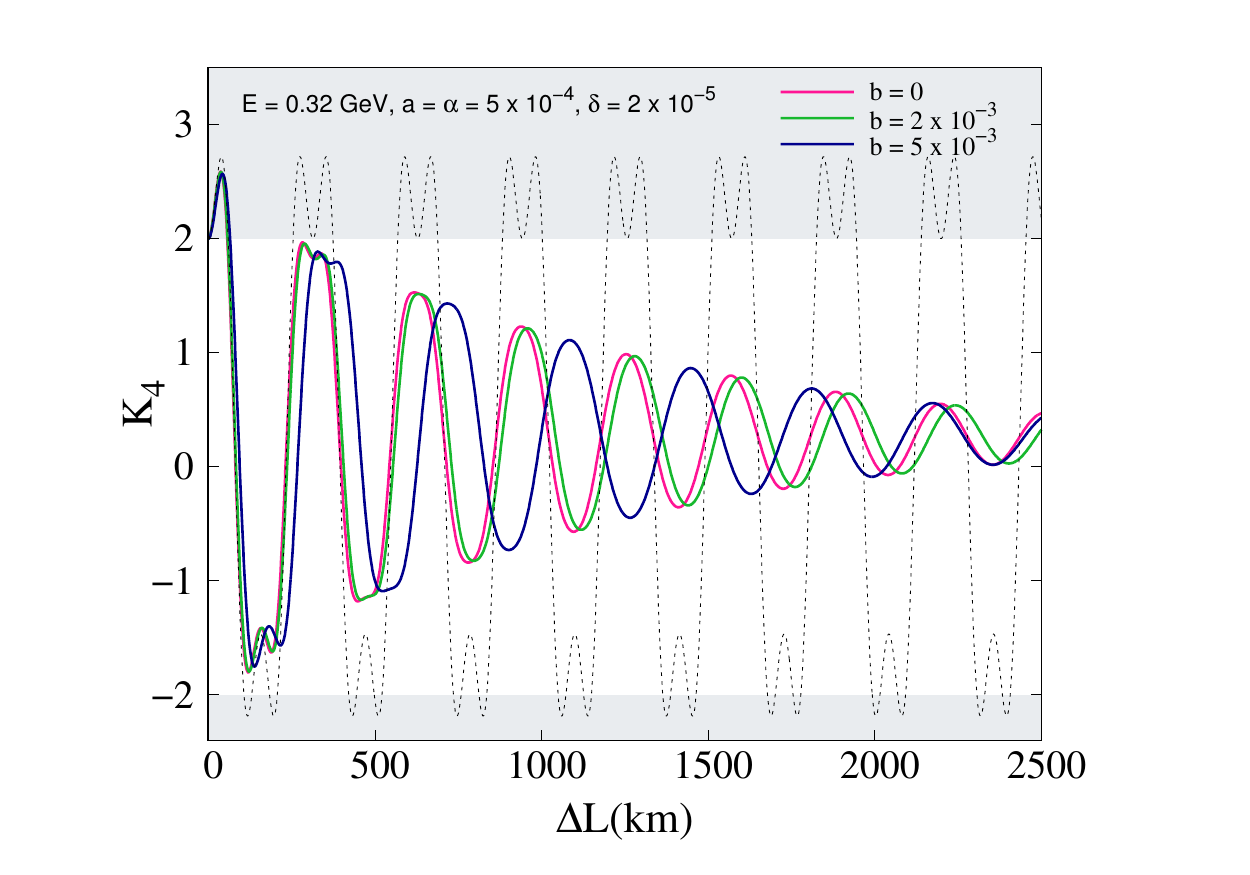}
\caption{\footnotesize{$K_4$ as a function of $\Delta L$ for  quantum decoherence (case 2 (b)). 
}}
\label{k4_decohb}
\end{figure}
In Fig.~\ref{k4_decoh}, $K_4$ plotted as a function of $\Delta L$ for  quantum decoherence (case 2 (a)). 
 As expected, the LGI parameter $K_4$ gradually decreases and falls within the classical limit.  It can be noted from the plot that 
{{ $K_4 \to 0.268$}} as we  go to large values of $\Delta L$.

\begin{figure}[htb!]
\centering
\includegraphics[width=0.65\linewidth]{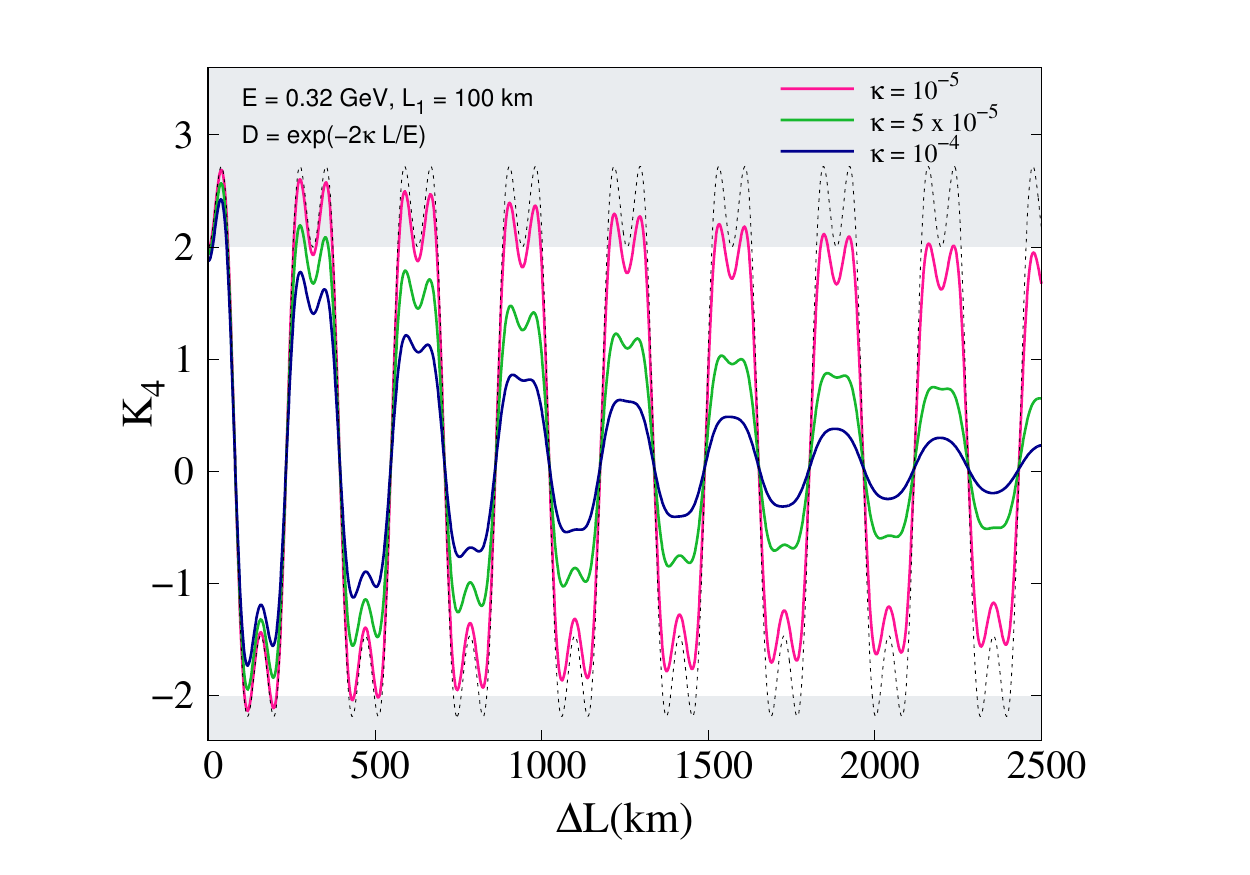}
\caption{\footnotesize{$K_4$ as a function of $\Delta L$ for  invisible neutrino decay (case 3). }}
\label{k4_decay}
\end{figure}  

 Let us now consider a case which violates energy conservation by taking $\beta=d=0$ and all other parameters to be non-zero (see Appendix A, case 5). 
For this case, $C_{12}$ is found to be
\begin{eqnarray}
    C_{12}&=&e^{-2\delta \Delta L} \cos^2{2\theta}+e^{-(a+\alpha) \Delta L}\sin^2{2\theta}\left[\cos (\Omega \Delta L)+\frac{(-a+\alpha)}{\Omega}\sin (\Omega \Delta L)\right]
\end{eqnarray}
And, $K_{4}$ can be found to be
\begin{eqnarray}
    K_{4}&=&3\left[ \exp\left(-2\delta \Delta L\right) \cos^2{2\theta}+\exp\left(-(a+\alpha) \Delta L\right)\sin^2{2\theta}\left[\cos (\Omega \Delta L)+\frac{(-a+\alpha)}{\Omega}\sin (\Omega \Delta L)\right]\right]
    \nonumber \\
     &-& 
    \exp\left(-6\delta \Delta L\right) \cos^2{2\theta}+\exp\left(-3(a+\alpha) \Delta L\right)\sin^2{2\theta}\left[\cos (3\Omega \Delta L)+\frac{(-a+\alpha)}{\Omega}\sin (3\Omega \Delta L)\right] \nonumber\\
\end{eqnarray}
In Fig.~\ref{k4_decohb}, $K_4$ is plotted against $\Delta L$ for this case. As expected, the LGI parameter $K_4$ gradually decreases and falls within the classical limit. 
We  can study the dependence of various decoherence parameters also from the four panels.  {{It can be noted from the plot that 
{{$K_4 \to 0.268$}} when $\delta = 0$ as we  go to large values of $\Delta L$. For other values of $\delta$, the limiting value is zero. }}
The third case is that of {\bf{Invisible neutrino decay}}. 
Using Eq.~\ref{eq:2}, we obtain  $C_{ij}$'s as
\begin{eqnarray}
C_{12} &=& \exp\left(\frac{-2 \kappa (L_{1}+ \Delta L)}{E}\right) \left(1 - 2 \sin^2 2\theta\sin^2 \left(\frac{\Delta m^2 \Delta L}{4 E}\right)\right)\nonumber\\
C_{23} &=& \exp\left(\frac{-2 \kappa (L_{1}+ 2\Delta L)}{E}\right) \left(1 - 2 \sin^2 2\theta\sin^2 \left(\frac{\Delta m^2 \Delta L}{4 E}\right)\right)\nonumber\\
C_{34} &=& \exp\left(\frac{-2 \kappa (L_{1}+ 3\Delta L)}{E}\right) \left(1 - 2 \sin^2 2\theta\sin^2 \left(\frac{\Delta m^2 \Delta L}{4 E}\right)\right)\nonumber\\
C_{14} &=& \exp\left(\frac{-2 \kappa (L_{1}+ 3\Delta L)}{E}\right) \left(1 - 2 \sin^2 2\theta\sin^2 \left(\frac{3\Delta m^2 \Delta L}{4 E}\right)\right)
\end{eqnarray}

\begin{figure}
\centering
\includegraphics[width=0.66\linewidth]{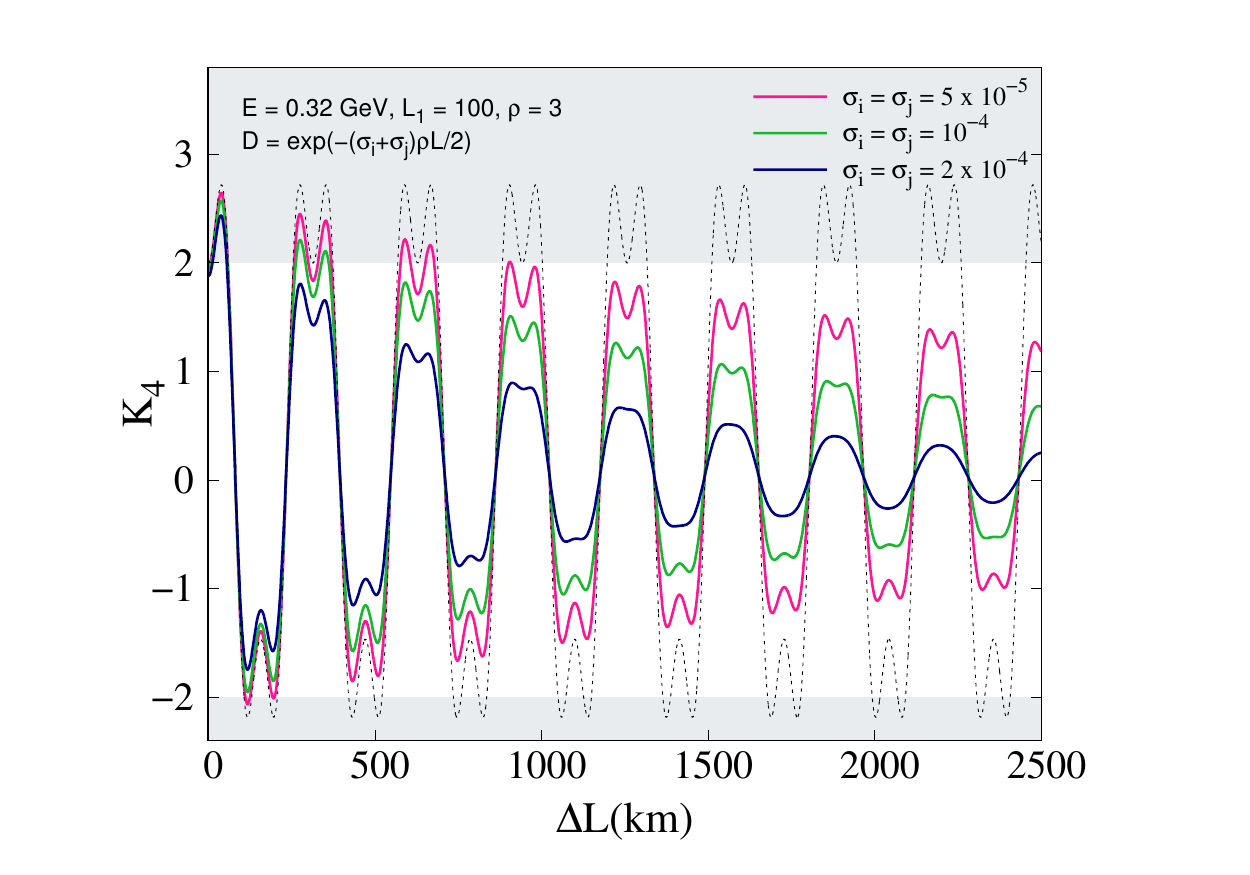}
\caption{\footnotesize{$K_4$ as a function of $\Delta L$ for the case of neutrino absorption (case 5). }}
\label{k4_absorb}
\end{figure}

which leads to the LGI parameter
\begin{eqnarray}
K_4 &=& \left[\exp\left(\frac{-2 \kappa (L_{1}+ \Delta L)}{E}\right)+ \exp\left(\frac{-2 \kappa (L_{1}+ 2\Delta L)}{E}\right) + \exp\left(\frac{-2 \kappa (L_{1}+ 3\Delta L)}{E}\right)\right] \nonumber\\
& \times & \left(1 - 2 \sin^2 2\theta\sin^2 \left(\frac{\Delta m^2 \Delta L}{4 E}\right)\right)- \exp\left(\frac{-2 \kappa (L_{1}+ 3\Delta L)}{E}\right) \nonumber\\ & \times & \left(1 - 2 \sin^2 2\theta\sin^2 \left(\frac{3\Delta m^2 \Delta L}{4 E}\right)\right)
\end{eqnarray}
 Fig.~\ref{k4_decay}, $K_4$ is plotted against $\Delta L$ for this case. {{It can be noted from the plot that 
{{$K_4 \to 0$}}  as we  go to large values of $\Delta L$.  }}
We now describe the case corresponding to {\bf{oscillation into sterile neutrinos}}. 
Here 
using Eq.~\ref{eq:3}, we get 
\begin{eqnarray}
    C_{12} &=& \exp\left(-\frac{\kappa  \left(L_1^2+\Delta L^2\right)}{4 E^2}\right) \left(1-2 \sin ^2(2 \theta ) \sin ^2\left(\frac{{\Delta m^2} (\Delta L)}{4 {E}}\right)\right)\nonumber \\
    C_{23} &=& \exp\left(-\frac{\kappa  \left(L_1^2+2\Delta L^2+2L_1 \Delta L \right)}{4 E^2}\right) \left(1-2 \sin ^2(2 \theta ) \sin ^2\left(\frac{{\Delta m^2} (\Delta L)}{4 {E}}\right)\right)\nonumber \\
    C_{34} &=& \exp\left(-\frac{\kappa  \left(L_1^2+5\Delta L^2+4L_1 \Delta L \right)}{4 E^2}\right) \left(1-2 \sin ^2(2 \theta ) \sin ^2\left(\frac{{\Delta m^2} (\Delta L)}{4 {E}}\right)\right)\nonumber \\
    C_{14} &=& \exp\left(-\frac{\kappa  \left(L_1^2+9\Delta L^2\right)}{4 E^2}\right) \left(1-2 \sin ^2(2 \theta ) \sin ^2\left(\frac{{\Delta m^2} (3\Delta L)}{4 {E}}\right)\right)
\end{eqnarray}
which leads to the LGI parameter,
\begin{eqnarray}
    K_4&=&\Bigg[\exp\left(\frac{-\kappa  \left(L_1^2+\Delta L^2\right)}{4 {E}^2}\right)
    +
    \exp\left(\frac{-\kappa  \left(L_1^2+2\Delta L^2+2L_1 \Delta L \right)}{4 {E}^2}\right)
    \nonumber\\
    &+&
    \exp\left(\frac{-\kappa  \left(L_1^2+5\Delta L^2+4L_1 \Delta L \right)}{4 {E}^2}\right)\Bigg]
  \times
   \left(1-2 \sin ^2(2 \theta ) \sin ^2\left(\frac{{\Delta m^2} {\Delta L}}{4 {E}}\right)\right) \nonumber\\
    &-&
    \exp\left(\frac{-\kappa  \left(L_1^2+9\Delta L^2 \right)}{4 {E}^2}\right) \times \left(1-2 \sin ^2(2 \theta ) \sin ^2\left(\frac{3\Delta m^2 \Delta L}{4 {E}}\right)\right)
\end{eqnarray}

\begin{figure}
\centering
\includegraphics[width=0.66\linewidth]{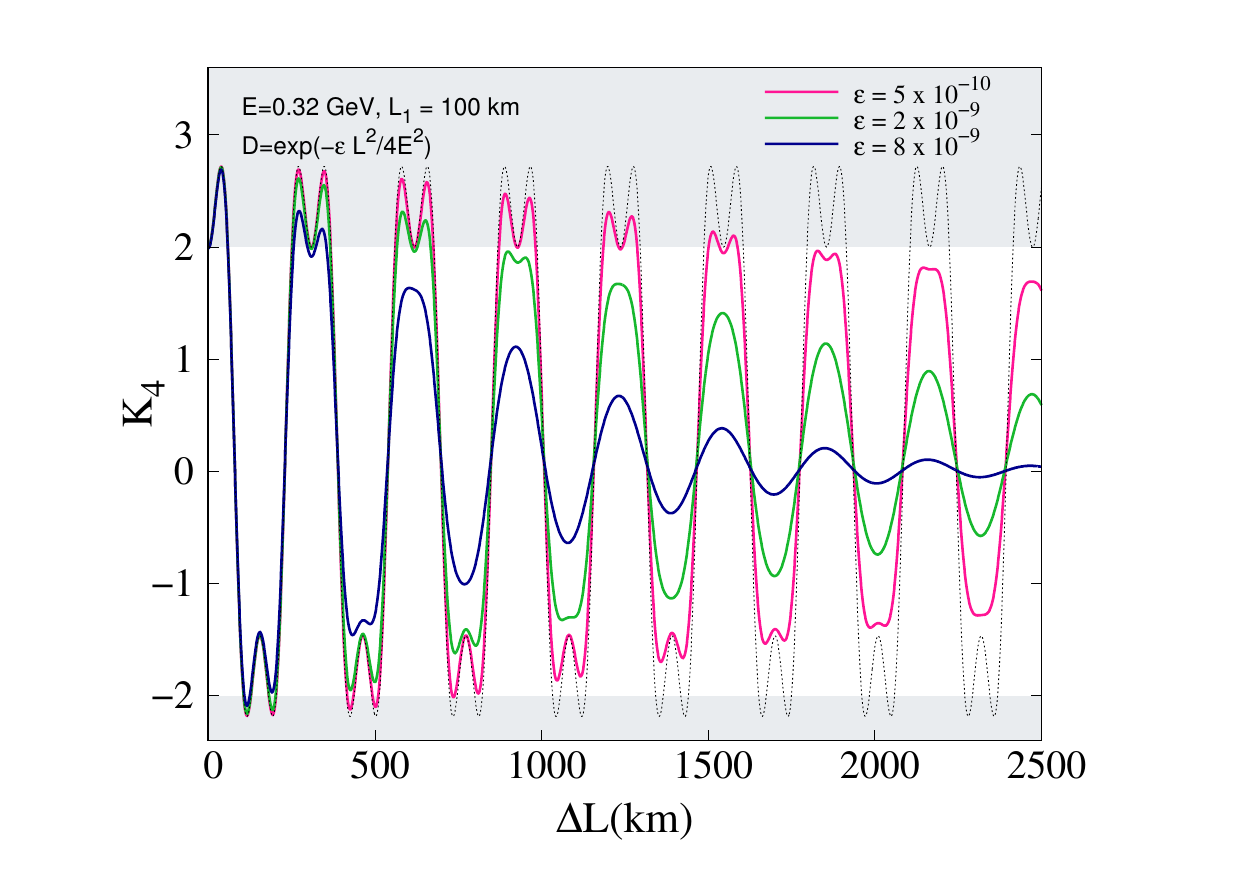}
\caption{\footnotesize{$K_4$ as a function of $\Delta L$ for  oscillation into sterile neutrino (case 4).  }}
\label{k4_sterile}
\end{figure} 
In Fig.~\ref{k4_sterile}, $K_4$ is plotted against $\Delta L$ for this case. {{It can be noted from the plot that 
{{$K_4 \to 0$}}   as we  go to large values of $\Delta L$. }}

Finally, we describe the case of {\bf{absorption of neutrinos}}. 
Using Eq.~\ref{eq:4}, the correlation functions are given by
\begin{eqnarray}
C_{12} &=& \exp [-\kappa \rho (L_{1}+ \Delta L)] \left(1 - 2 \sin^2 2\theta\sin^2 \left(\frac{\Delta m^2 \Delta L}{4 E}\right)\right) \nonumber\\
C_{23} &=& \exp [-\kappa \rho (L_{1}+ 2\Delta L)] \left(1 - 2 \sin^2 2\theta\sin^2 \left(\frac{\Delta m^2 \Delta L}{4 E}\right)\right)\nonumber\\
C_{34} &=& \exp[-\kappa \rho (L_{1}+ 3\Delta L)] \left(1 - 2 \sin^2 2\theta\sin^2 \left(\frac{\Delta m^2 \Delta L}{4 E}\right)\right)\nonumber\\
C_{14} &=& \exp[-\kappa \rho (L_{1}+ 3\Delta L)] \left(1 - 2 \sin^2 2\theta\sin^2 \left(\frac{3\Delta m^2 \Delta L}{4 E}\right)\right)
\end{eqnarray}
which leads to the LGI parameter
\begin{eqnarray}
K_4 &=& \left(
\exp[-\kappa \rho (L_{1}+ \Delta L)]+ \exp[- \kappa \rho (L_{1}+ 2\Delta L)] + \exp[-\kappa \rho (L_{1}+ 3\Delta L)]\right) \nonumber\\
&& \times  \left(1 - 2 \sin^2 2\theta\sin^2 \left(\frac{\Delta m^2 \Delta L}{4 E}\right)\right)- \exp[-\kappa \rho (L_{1}+ 3\Delta L)] \nonumber\\ && \times  \left(1 - 2 \sin^2 2\theta\sin^2 \left(\frac{3\Delta m^2 \Delta L}{4 E}\right)\right)
\end{eqnarray}
This case is shown in Fig.~\ref{k4_absorb}.   {{It can be noted from the plot that 
{{$K_4 \to 0$}}  as we  go to large values of $\Delta L$.  }}
 
\section{Conclusion}
\label{sec:5}
 
 Leggett-Garg inequalities have been studied in different physical systems~\cite{speakable,emary2013leggett}. Unlike other probes (such as photons) used to study LGI, neutrinos provide us with a unique opportunity as the mean free path of neutrinos is extremely large.  This  facilitates tests of LGI to be carried out over macroscopic distances. Indeed, in the experimental data, two experiments have reported violation of LGI at very high significance~\cite{Formaggio:2016cuh, Fu:2017hky}.

It is remarkable that neutrino oscillation experiments operating at different energy and length scales lend a very strong support to the standard mechanism of neutrino mass and mixing among the three active neutrino flavours. Historically, other mechanisms (such as decoherence) were considered as one of the plausible mechanisms for the flavour conversion of neutrinos. However, such scenarios can at best play a sub-dominant role in the physics of neutrino oscillations. Here,  we have explored  a class of non-standard oscillation effects which leave their imprints on the neutrino oscillation in the form of damping terms. 
In order to demonstrate the ideas in the present article, we have restricted ourselves to the case of two flavour neutrinos.

We have shown that the oscillation probability is affected in two ways in presence of such damping terms. Either the oscillatory term gets damped (cases 1 and 2) or the overall probability suffers leading to a loss of unitarity (cases 3, 4 and 5). As far as the LGI parameter, $K_4$ is concerned, $K_4$ also gets damped.  As a consequence, it falls within the classical regime ($-2 \leq K_4 \leq 2$) beyond a certain value  of $\Delta L$ for all the cases.  
The nature of this fall depends on the particular scenario.  The results for intrinsic wave packet decoherence (case 1) and quantum decoherence (case 2(a)) are very similar as expected~\cite{Ohlsson:2000mj}.  The stationarity condition is satisfied in these two cases.  However, the other three cases (3, 4 and 5) have peculiar outcome and the stationarity condition is not satisfied, \ie, $K_4$ depends not only on $\Delta L$ but also $L_1$ in these cases.

\section*{Acknowledgements}
The use of HPC cluster at SPS, JNU funded by DST-FIST is acknowledged. SS acknowledges financial support in the form of fellowship from University Grants Commission. PM would like to acknowledge funding from University Grants Commission under UPE II at JNU and Department of Science and Technology under DST- PURSE at JNU. The work of PM is partially supported by the European Union's Horizon 2020 research and innovation programme under the Marie Skodowska-Curie grant agreement No 690575 and 674896.

\appendix
\renewcommand\theequation{A\arabic{equation}}
\setcounter{equation}{0} 
\section*{Appendix A: Quantum decoherence effects in a bottom up approach}
\label{app}

We describe the evolution of neutrinos created in a state of definite flavour using the language of density matrices. Decoherence effects in neutrino propagation are introduced using the Liouville-Lindblad formalism for open quantum systems~\cite{Ellis1984381,Banks:1983by,Liu:1993ji,Benatti:2000ph}. According to it, the density matrix $\rho$ describing the neutrino flavour evolves according to 
\begin{equation}\label{EqLiouvilleLindbladDef}
 \frac{\partial\rho}{\partial t} = -i \left[\ham,\,\rho\right] + \mathcal{D}\left[\rho\right] ~
\end{equation}
where the Hamiltonian, $\ham$, is responsible for the usual unitary evolution, and the extra term, $\mathcal{D}\left[\rho\right]$, for non-unitary evolution, \ie, decoherence. A general parametrisation of $\mathcal{D}\left[\rho\right]$ is given in Ref.~\cite{Benatti:2000ph}.

\subsection*{Two generation case}

A commonly used form for ${\cal D} [\rho]$ was  given by Lindblad using quantum dynamical semi-groups~\cite{Lindblad:1975ef}. For a two-level quantum system, it is possible to expand all the operators in the $SU(2)$ Hermitian basis, i.e.,
\begin{eqnarray}
 \rho &=& \frac{1}{2} \left[p_\mu \lambda_\mu \right] = \frac{1}{2} \left[p_0 I +  p_i  \lambda_i \right] \nonumber\\
 H &=& \frac{1}{2} \left[h_\mu  \lambda_\mu \right]  = \frac{1}{2} \left[h_0 I +  h_i \lambda_i \right] \\
 {\cal D} [\rho] &=& \frac{1}{2} \left[ \lambda_\mu d_{\mu\nu}\rho_\nu \right] \nonumber
\end{eqnarray}
which leads to the equation of motion in the component form
\begin{equation}\label{equ:eomcomp}
 \dot p_\mu = (h_{\mu \nu} + d_{\mu\nu}) p_\nu ~
\end{equation}
where the subscripts $\mu,\nu$ depend upon the number of flavours. For two flavours, $\mu,\nu=0,1,2,3$ and for
three flavours, $\mu,\nu=0,\dots,8$. The matrix elements $h_{\mu\nu}$ are usually fixed from the form of the
Hamiltonian, while the elements of the matrix $d_{\mu\nu}$ can be fixed by assuming that the laws of
thermodynamics hold. If one imposes the requirement of the monotonic increase of von-Neumann entropy ($S=-{\rm {Tr}}[\rho \ln \rho]$), which leads to hermiticity of operators,
conservation of average value of energy, conservation of probability, etc., we can further constrain the elements
$d_{\mu\nu}$. For the two-flavour case, the basis is given by the three Pauli $SU(2)$ matrices $\{\sigma_i\}$ along with the identity matrix $I_2$.
We have the following matrix equation (from \equ{eomcomp}) with six allowed decoherence parameters as in Ref.~\cite{Hooper:2005jp}:
\begin{eqnarray}\label{equ:fullmateqn}
 \frac{d}{dt}\begin{pmatrix}
   p_0 \\
     p_1 \\
     p_2 \\
     p_3
 \end{pmatrix} &=& (-2)\left[
 \begin{pmatrix}
  0 & 0 & 0 & 0\\
  0  & a & b+\omega & d \\
  0 & b-\omega  & \alpha & \beta \\
  0& d & \beta &\delta
 \end{pmatrix} \right]\begin{pmatrix}
  p_0 \\
  p_1 \\
  p_2 \\
  p_3
 \end{pmatrix} ~
\end{eqnarray}
where $\omega = \Delta m ^2/\left(4E\right)$  with $\Delta m^2=m_i^2-m_j^2$ the standard oscillation term. $p_0 =1 $ by requiring ${\rm{Tr}}[\rho]=1$. We impose ${\rm{Tr}}[\rho]=1$ and ${\rm{Tr}}[\dot \rho] = 0$, which leads to conservation of probability and to the first row and column of the total matrix in the above equation being
zero. Trace is preserved during evolution and hence $p_0$ remains constant. The remaining three components of
$\rho(t)$ can be obtained by solving  the
  set of three coupled equations subject to the initial condition
\begin{eqnarray}\label{equ:rhoal}
 \rho_{\nu_\alpha} (0)  &=&
 \begin{pmatrix}
  U_{\alpha 1} ^2  & U_{\alpha 1} U_{\alpha 2} \\
  U_{\alpha 2} U_{\alpha 1}  &   U_{\alpha 2} ^2
 \end{pmatrix} ~
\end{eqnarray}
The neutrino flavour density matrix for the ``pure" flavour states $\ket{\nu_e} = (\cos \theta, \sin\theta)$ and
$\ket{\nu_\mu}=(-\sin \theta, \cos \theta)$ at the initial time $t=0$ are given by
\begin{eqnarray}
\rho_{\nu_e} (0) &=& \ket{\nu_e}\bra{\nu_e} = \begin{pmatrix}  \cos^2 \theta & \cos \theta \sin\theta \\\cos\theta \sin\theta & \sin^2\theta \end{pmatrix} \nonumber\\
\rho_{\nu_\mu} (0) &=& \ket{\nu_\mu}\bra{\nu_\mu} = \begin{pmatrix}  \sin^2 \theta & -\cos \theta
\sin\theta \\
-\cos\theta \sin\theta & \cos^2\theta \end{pmatrix}~ \label{equ:initialrho}
\end{eqnarray}
Now, we know that
\begin{eqnarray}
 \rho  (t)  &=& \frac{1}{2} \left[ p_0 I + \vec p (t)\cdot \vec \sigma \right] = \frac{1}{2} \begin{pmatrix}
 1 + p_3(t) & p_1(t) - i p_2 (t) \\
 p_1 (t) + i p_2 (t) & 1-p_3 (t)
 \end{pmatrix}~
 \label{equ:densityp}
\end{eqnarray}
where we have used the condition ${\rm{Tr}}[\rho]=1$ to get $p_0 = 1$. We readily find $p_i$ for the state
$\ket{\nu_e}$ as
\begin{equation}
 p_1 (0) = \sin 2\theta, \quad p_2 (0) = 0, \quad p_3 (0) = \cos 2\theta~ \label{equ:nuecomp}
\end{equation}
while for  $\ket{\nu_\mu}$ the components $p_i$ are
\begin{equation}
 p_1 (0) = -\sin 2 \theta, \quad p_2 (0) = 0, \quad p_3 (0) = -\cos 2\theta~ \label{equ:numucomp}
\end{equation}
Using \equs{nuecomp}{numucomp}, the initial pure density matrices (\equ{initialrho}) can be expressed as
\begin{eqnarray}
\rho_{\nu_e} (0)  &=&   \begin{pmatrix} \frac{1}{2} + \cos 2 \theta &  \sin 2 \theta \\
\sin 2 \theta & \frac{1}{2} - \cos 2\theta \end{pmatrix} \nonumber\\
\rho_{\nu_\mu} (0)  &=& \begin{pmatrix} \frac{1}{2} - \cos 2 \theta &  -\sin 2 \theta \\
-\sin 2 \theta & \frac{1}{2} + \cos 2\theta \end{pmatrix} = 1-\rho_{\nu_e} (0) ~
\end{eqnarray}
The diagonal elements of $\rho (t)$ are referred to as populations while the off-diagonal elements
as coherences. The phase information is contained in the coherences of the density matrix.
Let us define the $3 \times 3$ block in \equ{fullmateqn} connecting the components
$p_i~(i\neq 0)$ 
\begin{eqnarray}
 {\cal L} &=&
 \begin{pmatrix}
  a & b-\omega & d \\
  b + \omega & \alpha & \beta \\
  d &\beta & \delta
 \end{pmatrix}
\end{eqnarray}
and denote
\begin{equation}\label{equ:Mdef}
 {\cal M} = e^{-2 {\cal L} t}
\end{equation}
so that the solutions $p_i (t)$ of the differential equation
\begin{equation}\label{equ:diffeqpi}
 \dot{p}_i = \left(-2\right) \sum_{j=1}^3 \mathcal{L}_{ij} p_j
\end{equation}
can be written in terms of the elements of the exponentiated matrix ${\cal M}$,
\begin{eqnarray}\label{equ:eqspievol}
 p_1(t) &=&  {\cal M}_{11} p_1 (0) +  {\cal M}_{12}p_2 (0) +    {\cal M}_{13} p_3 (0)\nonumber \\
 p_2(t) &=&  {\cal M}_{21} p_1 (0) + {\cal M}_{22} p_2(0) +   {\cal M}_{23} p_3 (0)  \\
 p_3(t) &=&  {\cal M}_{31} p_1 (0) + {\cal M}_{32} p_2(0)  +  {\cal M}_{33} p_3(0)~ \nonumber
\label{equ:pasm}
\end{eqnarray}
where $p_{i} (0)$ are the components of the initial density matrix given in \equs{rhoal}{initialrho}.
Finally, the neutrino oscillation probability $P_{\alpha \beta}$ for the transition $\nu_\alpha \rightarrow \nu_\beta$ can be computed using
\begin{eqnarray}
P_{\alpha \beta} (t) &=& {\mathrm{Tr}} [\rho_{\nu_\alpha} (t) \, \rho_{\nu_\beta} (0)]~
\end{eqnarray}
where $\rho_{\nu_\beta} (0)$ is the ``pure" neutrino density matrix corresponding to flavour $\nu_\beta$ at $t=0$
and $\rho_{\nu_\alpha} (t)$ is the density matrix at $t$ for flavour $\nu_\alpha$. Hence, the probability of the transition $\nu_\alpha \rightarrow \nu_\mu$ is
\begin{eqnarray}
 P_{\alpha\mu} &=& {\mathrm{Tr}} [\rho_{\nu_\mu} (0) \, \rho (t)] \nonumber\\
               &=& {\mathrm{Tr}} \left[ \begin{pmatrix} \sin^2\theta &- \sin\theta \cos \theta \\ -\sin\theta \cos \theta&
                   \cos^2 \theta
                   \end{pmatrix} \frac{1}{2}\begin{pmatrix} 1+p_3 & p_1-ip_2 \\ p_1 +ip_2& 1- p_3
                   \end{pmatrix} \right]\nonumber\\
               &=& \frac{1}{2} \left[1 - p_3 (t) \cos(2\theta) - p_1 (t)\sin(2\theta) \right] ~
 \label{equ:pemu}
\end{eqnarray}
or, equivalenty,
\begin{eqnarray}
 P_{\alpha\mu}
 &=& \frac{1}{2} + \frac{1}{2} \left[- \cos^2(2\theta) {\cal M}_{33}  - \sin^2(2\theta)
 {\cal M}_{11} - \frac{1}{2}\sin (4\theta) ({\cal M}_{13}+{\cal M}_{31}) \right]~
 \label{equ:probasm}
\end{eqnarray}
 So, for different scenarios of quantum decoherence, we have distinct forms of ${\cal L}$ and then we can compute the
probability by finding the elements of ${\cal M}$ defined above. The probabilities in \equ{probasm} depend upon
time in general via the elements of ${\cal M}$ and they also depend upon the relative magnitude of $\omega$ with
respect to $a,b,d,\alpha,\beta,\delta$. In particular, if decoherence parameters are small compared to
$\omega$, we expect to see oscillations; else, the exponential damping terms will dominate the evolution. When $L$ is large (for astrophysical neutrinos)  which leads to loss of phase information and we can only access the
incoherent sum of probabilities.

\subsection*{Different cases of quantum decoherence}

To obtain a general expression for the probability in the two-flavour case, we
consider the following simple cases:

\begin{enumerate}
 \item
  \textbf{Standard oscillations:} all decoherence parameters ($a,b,d,\alpha,\beta,\delta$) are set to zero, \ie,
  \begin{eqnarray}
   {\cal L} &=&
   \begin{pmatrix}
    0  & -\omega  & 0 \\
    \omega & 0 & 0 \\
     0 & 0 & 0 
   \end{pmatrix}~
  \end{eqnarray}
  for which we get
  \begin{eqnarray}
  {\cal M}_{33} &=& 1 \nonumber\\
  {\cal M}_{11} &=& \cos(2 \omega t)  \\
  {\cal M}_{13} + {\cal M}_{31} &=& 0  \nonumber
  \end{eqnarray}
  and so
  \begin{eqnarray} 
   P_{e\mu}
   &=& \frac{1}{2} \left[ (1+ U_{e1}^2 - U_{e2}^2) U_{\mu 1}^2 +
   (1- U_{e1}^2 + U_{e2}^2) U_{\mu 2}^2 + 4 U_{e1} U_{e2} U_{\mu 1} U_{\mu 2} \cos{2 \omega t} \right]~.
  \end{eqnarray}
  \equ{probasm} leads to
  \begin{eqnarray}\label{equ:PStdNonZ}
   P_{e\mu}
   &=& \frac{1}{2} + \frac{1}{2} \left[  -\cos^2 (2\theta) - \sin^2 (2\theta) \cos ({ 2 \omega t} )
   \right]\nonumber\\
   &=& \frac{1}{2} \sin^2 (2\theta) [1-\cos (2 \omega t)]~
  \end{eqnarray}
  This matches the standard expression for probability for pure state evolution. Averaging over time, we get averaged oscillations,
  \begin{eqnarray}
   P_{e\mu} &\to& \sum_i |U_{e i}|^2 \, |U_{\mu i}|^2 = \frac{1}{2} \sin^2 (2 \theta)~
   \label{equ:asympso}
  \end{eqnarray}
  \item
   \textbf{Minimal decoherence scenario:} final row and column of ${\cal L}$ have zero-valued entries, \ie,
   $d=\beta=\delta=0$, so that
   \begin{eqnarray}
    {\cal L} &=&
    \begin{pmatrix} a  & b-\omega  & 0 \\ b+\omega & \alpha & 0
    \\ 0 & 0 & 0 \end{pmatrix}~
   \end{eqnarray}
   In particular, if we consider the form discussed in Ref.~\cite{Ohlsson:2000mj},
   \begin{eqnarray}
    (-2) {\cal L} &=&
    \begin{pmatrix} -d^2  & k  & 0 \\ -k & -d^2 & 0
    \\ 0 & 0 & 0 \end{pmatrix}~\end{eqnarray} we obtain
   \begin{eqnarray}
    {\cal M} &=&
    \begin{pmatrix} e^{-d^2 t} \cos (k t)  &e^{-d^2 t} \sin (k t)  & 0 \\ - e^{-d^2 t}\sin(kt) & e^{-d^2 t} \cos (k t) & 0
    \\ 0 & 0 & 1 \end{pmatrix} ~
   \end{eqnarray}
   from which
   \begin{eqnarray}
    {\cal M}_{33} &=& 1 \nonumber\\
    {\cal M}_{11} &=&  e^{-d^2 t} \cos (k t) \\
    {\cal M}_{13} + {\cal M}_{31} &=& 0 ~ \nonumber
   \end{eqnarray}
  \equ{probasm} gives
  \begin{eqnarray}
   P_{e\mu}
   &=& \frac{1}{2} + \frac{1}{2} \left[  -\cos^2 (2\theta) - \sin^2 (2\theta) e^{-d^2 t} \cos ({kt} )
   \right]\nonumber\\ &=& \frac{1}{2} \sin^2(2\theta) \left[ 1 -  e^{-d^2 t} \cos (k t)
   \right]~
   \label{equ:pemuecons}
  \end{eqnarray}
  For $k=2 \omega = \Delta m^2/2E$ and $t
  \simeq L$, this becomes
  \begin{eqnarray}\label{equ:PMinDecNonZ}
   P_{e\mu} &=& \frac{1}{2} \sin^2(2\theta) \left[ 1 -   e^{-d^2 L} \cos \left(\frac{\Delta m^2 }{2 E} L\right)
   \right]~
  \end{eqnarray}
  Averaging \equ{pemuecons} over time, we get the same limit as that
  for the case of no decoherence (\equ{asympso}) which depends upon the mixing angle $\theta$. Note that
  the  probability in the asymptotic limit is not the steady state value $(1/2)$ since this case corresponds to energy being conserved in the neutrino system.
  This case is the simplest possible extension of the standard oscillation probability with minimal decoherence
  parameters,  $a=\alpha = -d^2/(-2)$, $b=0$ and $\omega=-k/(-2)$.

  It is commonly assumed that the decoherence parameter $d^2$ has a power-law energy dependence of the form
  \begin{equation}\label{equ:Eqd2Def1}
   d^2 = \kappa E^n ~
  \end{equation}
  with $n$ an integer and $\kappa$ a constant with the right units (which depend on the value of $n$). Decoherence effects are expected to be highly suppressed, which will be reflected in a very small value of $\kappa$. Setting $n=0$ introduces energy-independent decoherence effects, while $n=2$ corresponds to a string-inspired decoherence model and $n=-1$, to Lorentz-invariant decoherence~\cite{Mavromatos:2007hv}. The condition for full decoherence, $2 d^2 L \gg 1$, can be satisfied for different combinations of $E$ and $L$ depending on the choice of $n$.
 \item
  \textbf{Decoherence scenario:} $\beta \neq 0$, so that
  \begin{eqnarray}
   {\cal L} &=&
   \begin{pmatrix} 0  & - \omega  & 0 \\ \omega & 0 & \beta
   \\ 0 & \beta & 0 \end{pmatrix}~
  \end{eqnarray}
  and so
  \begin{eqnarray}
   {\cal M}_{33} &=& \frac{-\omega^2 +\beta^2 \cos{(2 \Omega_\beta t)}}{-\Omega_\beta^2}
   \nonumber\\ {\cal M}_{11} &=&  \frac{\beta^2 -\omega^2 \cos{(2 \Omega_\beta t)}}{-\Omega_\beta^2} \\
   {\cal M}_{13} +{\cal M}_{31} &=& 0~ \nonumber
   \label{equ:mele3}
  \end{eqnarray}
  where $\Omega_\beta = \sqrt{\omega^2 - \beta^2}$. \equ{probasm} leads to
  \begin{eqnarray}
   P_{e\mu} 
   &=& \frac{1}{2} +\frac{1}{2}  \left[  \cos^2(2\theta)  \left\{ -\frac{\omega^2}{\Omega_\beta^2} +
   \frac{\beta^2}{\Omega_\beta^2} \cos({2  \Omega_{\beta} t} )
   \right\} + \sin^2(2\theta)
   \left\{  \frac{\beta^2}{\Omega_\beta^2} -
   \frac{\omega^2}{\Omega_\beta^2} \cos({2  \Omega_{\beta} t})
   \right\}   \right]~
   \nonumber \\
  \end{eqnarray}
  Averaging over time, we get
  \begin{eqnarray}
   P_{e\mu}  &\to& \frac{1}{2} +\frac{1}{2} \left\{-\frac{\omega^2}{\Omega^2_\beta} \cos^2(2\theta)
   +\frac{\beta^2}{\Omega_\beta^2} \sin^2(2\theta) \right\}~
   \label{equ:asympdecoh3}
  \end{eqnarray}
  \item
   \textbf{Decoherence scenario:} $d \neq 0$, so that
   \begin{eqnarray}
    {\cal L} &=&
    \begin{pmatrix} 0  & -\omega  & d \\ \omega & 0 & 0
    \\ d & 0 & 0 \end{pmatrix}~
   \end{eqnarray}
   and
   \begin{eqnarray}
    {\cal M}_{33} &=& \frac{\omega^2 - d^2 \cos{(2 \Omega_d t)}}{\Omega_d^2}
    \nonumber\\ {\cal M}_{11} &=&  \cos{(2 \Omega_d t)} \\
    {\cal M}_{13} + {\cal M}_{31} &=& \frac{-2d}{\Omega_d} \sin{(2 \Omega_d t)} ~ \nonumber \label{equ:mele4}
   \end{eqnarray}
   where $\Omega_d = \sqrt{\omega^2 - d^2}$.   Also,   this is the only case where we have ${\cal M}_{13} + {\cal
   M}_{31} \neq 0$. \equ{probasm} gives
   \begin{eqnarray}
    P_{e\mu}
    &=& \frac{1}{2} +  \frac{1}{2} \Bigg[\cos^2(2\theta)  \left\{  -\frac{\omega^2}{\Omega_d^2} +
    \frac{d^2}{\Omega_d^2} \cos({2   \Omega_{d} t})
    \right\} + \sin^2(2\theta)
    \left\{  -
    \cos({2 \Omega_{d} t})
    \right\} \nonumber \\
    && +  \sin(4\theta) \left\{ \frac{d}{\Omega_d} \sin(2  \Omega_d t)  \right\} \Bigg]~
   \end{eqnarray}
  Averaging over time, we get
   \begin{eqnarray}
    P_{e\mu}  &\to& \frac{1}{2}+ \frac{1}{2} \left\{-\frac{\omega^2}{\Omega^2_d} \cos^2(2\theta)
    \right\}~
    \label{equ:asympdecoh4}
   \end{eqnarray}
   Note that even though in cases 3 and 4 we have violation of conservation of energy in the neutrino system
   (since one of the elements in the final row or column in the ${\cal L}$ matrix is non-zero), the asymptotic
   limit is not exactly $1/2$ (steady-state value) (see \equs{asympdecoh3}{asympdecoh4}).
   This is because the terms of ${\cal M}$ are not exponentially suppressed, but are oscillatory functions (\equs{mele3}{mele4}).
 \item
  \textbf{Decoherence scenario:} $d = \beta = 0$, so that
  \begin{eqnarray}
   {\cal L} &=&
   \begin{pmatrix} a  & b-\omega  & 0 \\ b+\omega & \alpha & 0
   \\ 0 & 0 & \delta \end{pmatrix}~
  \end{eqnarray}
  Setting $\Omega_b = \sqrt{\omega^2 - b^2}$ and $a=\alpha$, we get
  \begin{eqnarray}
   {\cal M}_{33} &=& e^{-2 \delta t}
   \nonumber\\ {\cal M}_{11} &=&  e^{-2 a t} \cos{(2 \Omega_b t)} \\
   {\cal M}_{13} + {\cal M}_{31} &=& 0~
    \nonumber \label{equ:mele5}
  \end{eqnarray}
  We note from \equ{mele5} that ${\cal M}$ has either zero or exponentially
  suppressed entries. From \equ{probasm}, we have
  \begin{eqnarray}
   P_{e\mu}
   &=& \frac{1}{2} + \Big[  \cos^2(2\theta)  \left\{  - e^{-2   \delta t}
   \right\} + \sin^2(2\theta)
   \left\{  - e^{-2at} \cos({2  \Omega_{b} t})
   \right\}  \Big] ~
     \end{eqnarray}
  The asymptotic limit ($t \to \infty$) in this case is clearly
  \begin{eqnarray}
   P_{e\mu} &\to& \frac{1}{2}~\label{equ:asympdecoh}
  \end{eqnarray}
  So, the crucial requirement to get the steady-state value is that along with violation of conservation of energy in the neutrino system
  ($\delta \neq 0$), we must have some other decoherence parameters non-zero (such as $a$ or $\alpha \neq 0$).
  This leads to the exponentially suppressed elements of ${\cal M}$  which go to zero in the long $t$ limit. Even if $b =0$,
  we have the same steady-state limit. If $b=0$ and if we impose energy conservation ($\delta=0$),
  we recover case 2, for which the asymptotic limit is different (\equ{asympso}). This means non-conservation of
  energy is only a necessary but not a sufficient condition to reach the steady-state limit, $1/2$.
\end{enumerate}

\bibliography{references_2021}

\begin{thebibliography}{10}
\expandafter\ifx\csname bibnamefont\endcsname\relax
  \def\bibnamefont#1{#1}\fi
\expandafter\ifx\csname bibfnamefont\endcsname\relax
  \def\bibfnamefont#1{#1}\fi
\expandafter\ifx\csname url\endcsname\relax
  \def\url#1{\texttt{#1}}\fi
\expandafter\ifx\csname urlprefix\endcsname\relax\def\urlprefix{URL }\fi
\providecommand{\bibinfo}[2]{#2}
\providecommand{\eprint}[2][]{\url{#2}}

\bibitem{deSalas:2020pgw}
\bibinfo{author}{\bibfnamefont{P.~F.} \bibnamefont{de~Salas}},
  \bibinfo{author}{\bibfnamefont{D.~V.} \bibnamefont{Forero}},
  \bibinfo{author}{\bibfnamefont{S.}~\bibnamefont{Gariazzo}},
  \bibinfo{author}{\bibfnamefont{P.}~\bibnamefont{Mart\'\i{}nez-Mirav\'e}},
  \bibinfo{author}{\bibfnamefont{O.}~\bibnamefont{Mena}},
  \bibinfo{author}{\bibfnamefont{C.~A.} \bibnamefont{Ternes}},
  \bibinfo{author}{\bibfnamefont{M.}~\bibnamefont{T\'ortola}},
  \bibnamefont{and} \bibinfo{author}{\bibfnamefont{J.~W.~F.}
  \bibnamefont{Valle}}, \bibinfo{journal}{JHEP} \textbf{\bibinfo{volume}{02}},
  \bibinfo{pages}{071} (\bibinfo{year}{2021}), \eprint{2006.11237}.

\bibitem{T2K:2019bcf}
\bibinfo{author}{\bibfnamefont{K.}~\bibnamefont{Abe}} \emph{et~al.}
  (\bibinfo{collaboration}{T2K}), \bibinfo{journal}{Nature}
  \textbf{\bibinfo{volume}{580}}(\bibinfo{number}{7803}), \bibinfo{pages}{339}
  (\bibinfo{year}{2020}), \bibinfo{note}{[Erratum: Nature 583, E16 (2020)]},
  \eprint{1910.03887}.

\bibitem{NOvA:2021nfi}
\bibinfo{author}{\bibfnamefont{M.~A.} \bibnamefont{Acero}} \emph{et~al.}
  (\bibinfo{collaboration}{NOvA}) \eprint{2108.08219}.

\bibitem{LSND:2001aii}
\bibinfo{author}{\bibfnamefont{A.}~\bibnamefont{Aguilar-Arevalo}} \emph{et~al.}
  (\bibinfo{collaboration}{LSND}), \bibinfo{journal}{Phys. Rev. D}
  \textbf{\bibinfo{volume}{64}}, \bibinfo{pages}{112007}
  (\bibinfo{year}{2001}), \eprint{hep-ex/0104049}.

\bibitem{MiniBooNE:2018esg}
\bibinfo{author}{\bibfnamefont{A.~A.} \bibnamefont{Aguilar-Arevalo}}
  \emph{et~al.} (\bibinfo{collaboration}{MiniBooNE}), \bibinfo{journal}{Phys.
  Rev. Lett.} \textbf{\bibinfo{volume}{121}}(\bibinfo{number}{22}),
  \bibinfo{pages}{221801} (\bibinfo{year}{2018}), \eprint{1805.12028}.

\bibitem{MiniBooNE:2020pnu}
\bibinfo{author}{\bibfnamefont{A.~A.} \bibnamefont{Aguilar-Arevalo}}
  \emph{et~al.} (\bibinfo{collaboration}{MiniBooNE}), \bibinfo{journal}{Phys.
  Rev. D} \textbf{\bibinfo{volume}{103}}(\bibinfo{number}{5}),
  \bibinfo{pages}{052002} (\bibinfo{year}{2021}), \eprint{2006.16883}.

\bibitem{Mueller:2011nm}
\bibinfo{author}{\bibfnamefont{T.~A.} \bibnamefont{Mueller}} \emph{et~al.},
  \bibinfo{journal}{Phys. Rev. C} \textbf{\bibinfo{volume}{83}},
  \bibinfo{pages}{054615} (\bibinfo{year}{2011}), \eprint{1101.2663}.

\bibitem{MicroBooNE:2021rmx}
\bibinfo{author}{\bibfnamefont{P.}~\bibnamefont{Abratenko}} \emph{et~al.}
  (\bibinfo{collaboration}{MicroBooNE}) \eprint{arXiv:2110.14054 [hep-ex]}.

\bibitem{MicroBooNE:2021sne}
\bibinfo{author}{\bibfnamefont{P.}~\bibnamefont{Abratenko}} \emph{et~al.}
  (\bibinfo{collaboration}{MicroBooNE}) \eprint{arXiv:2110.14065 [hep-ex]}.

\bibitem{MicroBooNE:2021jwr}
\bibinfo{author}{\bibfnamefont{P.}~\bibnamefont{Abratenko}} \emph{et~al.}
  (\bibinfo{collaboration}{MicroBooNE}) \eprint{arXiv:2110.14080 [hep-ex]}.

\bibitem{Glashow:1961tr}
\bibinfo{author}{\bibfnamefont{S.~L.} \bibnamefont{Glashow}},
  \bibinfo{journal}{Nucl. Phys.} \textbf{\bibinfo{volume}{22}},
  \bibinfo{pages}{579} (\bibinfo{year}{1961}).

\bibitem{Weinberg:1967tq}
\bibinfo{author}{\bibfnamefont{S.}~\bibnamefont{Weinberg}},
  \bibinfo{journal}{Phys. Rev. Lett.} \textbf{\bibinfo{volume}{19}},
  \bibinfo{pages}{1264} (\bibinfo{year}{1967}).

\bibitem{osti_4767615}
\bibinfo{author}{\bibfnamefont{A.}~\bibnamefont{Salam}}, \bibinfo{journal}{pp
  367-77 of Elementary Particle Theory. Svartholm, Nils (ed.). New York, John
  Wiley and Sons, Inc., 1968.}
  \urlprefix\url{https://www.osti.gov/biblio/4767615}.

\bibitem{Eliezer:1975ja}
\bibinfo{author}{\bibfnamefont{S.}~\bibnamefont{Eliezer}} \bibnamefont{and}
  \bibinfo{author}{\bibfnamefont{A.~R.} \bibnamefont{Swift}},
  \bibinfo{journal}{Nucl. Phys. B} \textbf{\bibinfo{volume}{105}},
  \bibinfo{pages}{45} (\bibinfo{year}{1976}).

\bibitem{Fritzsch:1975rz}
\bibinfo{author}{\bibfnamefont{H.}~\bibnamefont{Fritzsch}} \bibnamefont{and}
  \bibinfo{author}{\bibfnamefont{P.}~\bibnamefont{Minkowski}},
  \bibinfo{journal}{Phys. Lett. B} \textbf{\bibinfo{volume}{62}},
  \bibinfo{pages}{72} (\bibinfo{year}{1976}).

\bibitem{Nussinov:1976uw}
\bibinfo{author}{\bibfnamefont{S.}~\bibnamefont{Nussinov}},
  \bibinfo{journal}{Phys. Lett. B} \textbf{\bibinfo{volume}{63}},
  \bibinfo{pages}{201} (\bibinfo{year}{1976}).

\bibitem{Bilenky:1978nj}
\bibinfo{author}{\bibfnamefont{S.~M.} \bibnamefont{Bilenky}} \bibnamefont{and}
  \bibinfo{author}{\bibfnamefont{B.}~\bibnamefont{Pontecorvo}},
  \bibinfo{journal}{Phys. Rept.} \textbf{\bibinfo{volume}{41}},
  \bibinfo{pages}{225} (\bibinfo{year}{1978}).

\bibitem{Kayser:1981ye}
\bibinfo{author}{\bibfnamefont{B.}~\bibnamefont{Kayser}},
  \bibinfo{journal}{Phys. Rev. D} \textbf{\bibinfo{volume}{24}},
  \bibinfo{pages}{110} (\bibinfo{year}{1981}).

\bibitem{Giunti:1991ca}
\bibinfo{author}{\bibfnamefont{C.}~\bibnamefont{Giunti}},
  \bibinfo{author}{\bibfnamefont{C.~W.} \bibnamefont{Kim}}, \bibnamefont{and}
  \bibinfo{author}{\bibfnamefont{U.~W.} \bibnamefont{Lee}},
  \bibinfo{journal}{Phys. Rev. D} \textbf{\bibinfo{volume}{44}},
  \bibinfo{pages}{3635} (\bibinfo{year}{1991}).

\bibitem{Kiers:1995zj}
\bibinfo{author}{\bibfnamefont{K.}~\bibnamefont{Kiers}},
  \bibinfo{author}{\bibfnamefont{S.}~\bibnamefont{Nussinov}}, \bibnamefont{and}
  \bibinfo{author}{\bibfnamefont{N.}~\bibnamefont{Weiss}},
  \bibinfo{journal}{Phys. Rev. D} \textbf{\bibinfo{volume}{53}},
  \bibinfo{pages}{537} (\bibinfo{year}{1996}), \eprint{hep-ph/9506271}.

\bibitem{Kiers:1997pe}
\bibinfo{author}{\bibfnamefont{K.}~\bibnamefont{Kiers}} \bibnamefont{and}
  \bibinfo{author}{\bibfnamefont{N.}~\bibnamefont{Weiss}},
  \bibinfo{journal}{Phys. Rev. D} \textbf{\bibinfo{volume}{57}},
  \bibinfo{pages}{3091} (\bibinfo{year}{1998}), \eprint{hep-ph/9710289}.

\bibitem{Giunti:1997wq}
\bibinfo{author}{\bibfnamefont{C.}~\bibnamefont{Giunti}} \bibnamefont{and}
  \bibinfo{author}{\bibfnamefont{C.~W.} \bibnamefont{Kim}},
  \bibinfo{journal}{Phys. Rev. D} \textbf{\bibinfo{volume}{58}},
  \bibinfo{pages}{017301} (\bibinfo{year}{1998}), \eprint{hep-ph/9711363}.

\bibitem{Grimus:1998uh}
\bibinfo{author}{\bibfnamefont{W.}~\bibnamefont{Grimus}},
  \bibinfo{author}{\bibfnamefont{P.}~\bibnamefont{Stockinger}},
  \bibnamefont{and} \bibinfo{author}{\bibfnamefont{S.}~\bibnamefont{Mohanty}},
  \bibinfo{journal}{Phys. Rev. D} \textbf{\bibinfo{volume}{59}},
  \bibinfo{pages}{013011} (\bibinfo{year}{1999}), \eprint{hep-ph/9807442}.

\bibitem{Cardall:1999ze}
\bibinfo{author}{\bibfnamefont{C.~Y.} \bibnamefont{Cardall}},
  \bibinfo{journal}{Phys. Rev. D} \textbf{\bibinfo{volume}{61}},
  \bibinfo{pages}{073006} (\bibinfo{year}{2000}), \eprint{hep-ph/9909332}.

\bibitem{Akhmedov:2009rb}
\bibinfo{author}{\bibfnamefont{E.~K.} \bibnamefont{Akhmedov}} \bibnamefont{and}
  \bibinfo{author}{\bibfnamefont{A.~Y.} \bibnamefont{Smirnov}},
  \bibinfo{journal}{Phys. Atom. Nucl.} \textbf{\bibinfo{volume}{72}},
  \bibinfo{pages}{1363} (\bibinfo{year}{2009}), \eprint{0905.1903}.

\bibitem{Kayser:2010pr}
\bibinfo{author}{\bibfnamefont{B.}~\bibnamefont{Kayser}} \bibnamefont{and}
  \bibinfo{author}{\bibfnamefont{J.}~\bibnamefont{Kopp}}
  \eprint{arXiv:1005.4081 [hep-ph]}.

\bibitem{Naumov_2010}
\bibinfo{author}{\bibfnamefont{D.~V.} \bibnamefont{Naumov}} \bibnamefont{and}
  \bibinfo{author}{\bibfnamefont{V.~A.} \bibnamefont{Naumov}},
  \bibinfo{journal}{Journal of Physics G: Nuclear and Particle Physics}
  \textbf{\bibinfo{volume}{37}}(\bibinfo{number}{10}), \bibinfo{pages}{105014}
  (\bibinfo{year}{2010}),
  \urlprefix\url{https://doi.org/10.1088/0954-3899/37/10/105014}.

\bibitem{Akhmedov:2010ms}
\bibinfo{author}{\bibfnamefont{E.~K.} \bibnamefont{Akhmedov}} \bibnamefont{and}
  \bibinfo{author}{\bibfnamefont{J.}~\bibnamefont{Kopp}},
  \bibinfo{journal}{JHEP} \textbf{\bibinfo{volume}{04}}, \bibinfo{pages}{008}
  (\bibinfo{year}{2010}), \bibinfo{note}{[Erratum: JHEP 10, 052 (2013)]},
  \eprint{1001.4815}.

\bibitem{Akhmedov2012}
\bibinfo{author}{\bibfnamefont{E.~K.} \bibnamefont{Akhmedov}},
  \bibinfo{author}{\bibfnamefont{D.}~\bibnamefont{Hernandez}},
  \bibnamefont{and} \bibinfo{author}{\bibfnamefont{A.~Y.}
  \bibnamefont{Smirnov}}, \bibinfo{journal}{Journal of High Energy Physics}
  \textbf{\bibinfo{volume}{2012}}(\bibinfo{number}{4}), \bibinfo{pages}{52}
  (\bibinfo{year}{2012}), ISSN \bibinfo{issn}{1029-8479},
  \urlprefix\url{https://doi.org/10.1007/JHEP04(2012)052}.

\bibitem{Egorov:2019vqv}
\bibinfo{author}{\bibfnamefont{V.~O.} \bibnamefont{Egorov}} \bibnamefont{and}
  \bibinfo{author}{\bibfnamefont{I.~P.} \bibnamefont{Volobuev}},
  \bibinfo{journal}{Phys. Rev. D}
  \textbf{\bibinfo{volume}{100}}(\bibinfo{number}{3}), \bibinfo{pages}{033004}
  (\bibinfo{year}{2019}), \eprint{1902.03602}.

\bibitem{Blennow:2005yk}
\bibinfo{author}{\bibfnamefont{M.}~\bibnamefont{Blennow}},
  \bibinfo{author}{\bibfnamefont{T.}~\bibnamefont{Ohlsson}}, \bibnamefont{and}
  \bibinfo{author}{\bibfnamefont{W.}~\bibnamefont{Winter}},
  \bibinfo{journal}{JHEP} \textbf{\bibinfo{volume}{06}}, \bibinfo{pages}{049}
  (\bibinfo{year}{2005}), \eprint{hep-ph/0502147}.

\bibitem{Bahcall:1972my}
\bibinfo{author}{\bibfnamefont{J.~N.} \bibnamefont{Bahcall}},
  \bibinfo{author}{\bibfnamefont{N.}~\bibnamefont{Cabibbo}}, \bibnamefont{and}
  \bibinfo{author}{\bibfnamefont{A.}~\bibnamefont{Yahil}},
  \bibinfo{journal}{Phys. Rev. Lett.} \textbf{\bibinfo{volume}{28}},
  \bibinfo{pages}{316} (\bibinfo{year}{1972}).

\bibitem{Barger:1981vd}
\bibinfo{author}{\bibfnamefont{V.~D.} \bibnamefont{Barger}},
  \bibinfo{author}{\bibfnamefont{W.-Y.} \bibnamefont{Keung}}, \bibnamefont{and}
  \bibinfo{author}{\bibfnamefont{S.}~\bibnamefont{Pakvasa}},
  \bibinfo{journal}{Phys. Rev. D} \textbf{\bibinfo{volume}{25}},
  \bibinfo{pages}{907} (\bibinfo{year}{1982}).

\bibitem{Valle:1983ua}
\bibinfo{author}{\bibfnamefont{J.~W.~F.} \bibnamefont{Valle}},
  \bibinfo{journal}{Phys. Lett. B} \textbf{\bibinfo{volume}{131}},
  \bibinfo{pages}{87} (\bibinfo{year}{1983}).

\bibitem{Barger:1998xk}
\bibinfo{author}{\bibfnamefont{V.~D.} \bibnamefont{Barger}},
  \bibinfo{author}{\bibfnamefont{J.~G.} \bibnamefont{Learned}},
  \bibinfo{author}{\bibfnamefont{S.}~\bibnamefont{Pakvasa}}, \bibnamefont{and}
  \bibinfo{author}{\bibfnamefont{T.~J.} \bibnamefont{Weiler}},
  \bibinfo{journal}{Phys. Rev. Lett.} \textbf{\bibinfo{volume}{82}},
  \bibinfo{pages}{2640} (\bibinfo{year}{1999}), \eprint{astro-ph/9810121}.

\bibitem{Pakvasa:1999ta}
\bibinfo{author}{\bibfnamefont{S.}~\bibnamefont{Pakvasa}},
  \bibinfo{journal}{AIP Conf. Proc.}
  \textbf{\bibinfo{volume}{542}}(\bibinfo{number}{1}), \bibinfo{pages}{99}
  (\bibinfo{year}{2000}), \eprint{hep-ph/0004077}.

\bibitem{Barger:1999bg}
\bibinfo{author}{\bibfnamefont{V.~D.} \bibnamefont{Barger}},
  \bibinfo{author}{\bibfnamefont{J.~G.} \bibnamefont{Learned}},
  \bibinfo{author}{\bibfnamefont{P.}~\bibnamefont{Lipari}},
  \bibinfo{author}{\bibfnamefont{M.}~\bibnamefont{Lusignoli}},
  \bibinfo{author}{\bibfnamefont{S.}~\bibnamefont{Pakvasa}}, \bibnamefont{and}
  \bibinfo{author}{\bibfnamefont{T.~J.} \bibnamefont{Weiler}},
  \bibinfo{journal}{Phys. Lett. B} \textbf{\bibinfo{volume}{462}},
  \bibinfo{pages}{109} (\bibinfo{year}{1999}), \eprint{hep-ph/9907421}.

\bibitem{Lindner:2001fx}
\bibinfo{author}{\bibfnamefont{M.}~\bibnamefont{Lindner}},
  \bibinfo{author}{\bibfnamefont{T.}~\bibnamefont{Ohlsson}}, \bibnamefont{and}
  \bibinfo{author}{\bibfnamefont{W.}~\bibnamefont{Winter}},
  \bibinfo{journal}{Nucl. Phys. B} \textbf{\bibinfo{volume}{607}},
  \bibinfo{pages}{326} (\bibinfo{year}{2001}), \eprint{hep-ph/0103170}.

\bibitem{DeRujula:1983ya}
\bibinfo{author}{\bibfnamefont{A.}~\bibnamefont{De~Rujula}},
  \bibinfo{author}{\bibfnamefont{S.~L.} \bibnamefont{Glashow}},
  \bibinfo{author}{\bibfnamefont{R.~R.} \bibnamefont{Wilson}},
  \bibnamefont{and} \bibinfo{author}{\bibfnamefont{G.}~\bibnamefont{Charpak}},
  \bibinfo{journal}{Phys. Rept.} \textbf{\bibinfo{volume}{99}},
  \bibinfo{pages}{341} (\bibinfo{year}{1983}).

\bibitem{Strumia:2002fw}
\bibinfo{author}{\bibfnamefont{A.}~\bibnamefont{Strumia}},
  \bibinfo{journal}{Phys. Lett. B} \textbf{\bibinfo{volume}{539}},
  \bibinfo{pages}{91} (\bibinfo{year}{2002}), \eprint{hep-ph/0201134}.

\bibitem{Maltoni:2004ei}
\bibinfo{author}{\bibfnamefont{M.}~\bibnamefont{Maltoni}},
  \bibinfo{author}{\bibfnamefont{T.}~\bibnamefont{Schwetz}},
  \bibinfo{author}{\bibfnamefont{M.~A.} \bibnamefont{Tortola}},
  \bibnamefont{and} \bibinfo{author}{\bibfnamefont{J.~W.~F.}
  \bibnamefont{Valle}}, \bibinfo{journal}{New J. Phys.}
  \textbf{\bibinfo{volume}{6}}, \bibinfo{pages}{122} (\bibinfo{year}{2004}),
  \eprint{hep-ph/0405172}.

\bibitem{Gago:2000qc}
\bibinfo{author}{\bibfnamefont{A.~M.} \bibnamefont{Gago}},
  \bibinfo{author}{\bibfnamefont{E.~M.} \bibnamefont{Santos}},
  \bibinfo{author}{\bibfnamefont{W.~J.~C.} \bibnamefont{Teves}},
  \bibnamefont{and}
  \bibinfo{author}{\bibfnamefont{R.}~\bibnamefont{Zukanovich~Funchal}},
  \bibinfo{journal}{Phys. Rev. D} \textbf{\bibinfo{volume}{63}},
  \bibinfo{pages}{073001} (\bibinfo{year}{2001}), \eprint{hep-ph/0009222}.

\bibitem{Ohlsson:2000mj}
\bibinfo{author}{\bibfnamefont{T.}~\bibnamefont{Ohlsson}},
  \bibinfo{journal}{Phys. Lett. B} \textbf{\bibinfo{volume}{502}},
  \bibinfo{pages}{159} (\bibinfo{year}{2001}), \eprint{hep-ph/0012272}.

\bibitem{Morgan:2004vv}
\bibinfo{author}{\bibfnamefont{D.}~\bibnamefont{Morgan}},
  \bibinfo{author}{\bibfnamefont{E.}~\bibnamefont{Winstanley}},
  \bibinfo{author}{\bibfnamefont{J.}~\bibnamefont{Brunner}}, \bibnamefont{and}
  \bibinfo{author}{\bibfnamefont{L.~F.} \bibnamefont{Thompson}},
  \bibinfo{journal}{Astropart. Phys.} \textbf{\bibinfo{volume}{25}},
  \bibinfo{pages}{311} (\bibinfo{year}{2006}), \eprint{astro-ph/0412618}.

\bibitem{Hooper:2005jp}
\bibinfo{author}{\bibfnamefont{D.}~\bibnamefont{Hooper}},
  \bibinfo{author}{\bibfnamefont{D.}~\bibnamefont{Morgan}}, \bibnamefont{and}
  \bibinfo{author}{\bibfnamefont{E.}~\bibnamefont{Winstanley}},
  \bibinfo{journal}{Phys. Rev. D} \textbf{\bibinfo{volume}{72}},
  \bibinfo{pages}{065009} (\bibinfo{year}{2005}), \eprint{hep-ph/0506091}.

\bibitem{Anchordoqui:2005gj}
\bibinfo{author}{\bibfnamefont{L.~A.} \bibnamefont{Anchordoqui}},
  \bibinfo{author}{\bibfnamefont{H.}~\bibnamefont{Goldberg}},
  \bibinfo{author}{\bibfnamefont{M.~C.} \bibnamefont{Gonzalez-Garcia}},
  \bibinfo{author}{\bibfnamefont{F.}~\bibnamefont{Halzen}},
  \bibinfo{author}{\bibfnamefont{D.}~\bibnamefont{Hooper}},
  \bibinfo{author}{\bibfnamefont{S.}~\bibnamefont{Sarkar}}, \bibnamefont{and}
  \bibinfo{author}{\bibfnamefont{T.~J.} \bibnamefont{Weiler}},
  \bibinfo{journal}{Phys. Rev. D} \textbf{\bibinfo{volume}{72}},
  \bibinfo{pages}{065019} (\bibinfo{year}{2005}), \eprint{hep-ph/0506168}.

\bibitem{Farzan:2008eg}
\bibinfo{author}{\bibfnamefont{Y.}~\bibnamefont{Farzan}} \bibnamefont{and}
  \bibinfo{author}{\bibfnamefont{A.~{\relax Yu}.} \bibnamefont{Smirnov}},
  \bibinfo{journal}{Nucl. Phys.} \textbf{\bibinfo{volume}{B805}},
  \bibinfo{pages}{356} (\bibinfo{year}{2008}).

\bibitem{Farzan:2008zv}
\bibinfo{author}{\bibfnamefont{Y.}~\bibnamefont{Farzan}},
  \bibinfo{author}{\bibfnamefont{T.}~\bibnamefont{Schwetz}}, \bibnamefont{and}
  \bibinfo{author}{\bibfnamefont{A.~Y.} \bibnamefont{Smirnov}},
  \bibinfo{journal}{JHEP} \textbf{\bibinfo{volume}{07}}, \bibinfo{pages}{067}
  (\bibinfo{year}{2008}), \eprint{0805.2098}.

\bibitem{Mehta:2011qb}
\bibinfo{author}{\bibfnamefont{P.}~\bibnamefont{Mehta}} \bibnamefont{and}
  \bibinfo{author}{\bibfnamefont{W.}~\bibnamefont{Winter}},
  \bibinfo{journal}{JCAP} \textbf{\bibinfo{volume}{1103}}, \bibinfo{pages}{041}
  (\bibinfo{year}{2011}), \eprint{1101.2673}.

\bibitem{Kersten:2015kio}
\bibinfo{author}{\bibfnamefont{J.}~\bibnamefont{Kersten}} \bibnamefont{and}
  \bibinfo{author}{\bibfnamefont{A.~Y.} \bibnamefont{Smirnov}},
  \bibinfo{journal}{Eur. Phys. J. C}
  \textbf{\bibinfo{volume}{76}}(\bibinfo{number}{6}), \bibinfo{pages}{339}
  (\bibinfo{year}{2016}), \eprint{1512.09068}.

\bibitem{Rasmussen:2017ert}
\bibinfo{author}{\bibfnamefont{R.~W.} \bibnamefont{Rasmussen}},
  \bibinfo{author}{\bibfnamefont{L.}~\bibnamefont{Lechner}},
  \bibinfo{author}{\bibfnamefont{M.}~\bibnamefont{Ackermann}},
  \bibinfo{author}{\bibfnamefont{M.}~\bibnamefont{Kowalski}}, \bibnamefont{and}
  \bibinfo{author}{\bibfnamefont{W.}~\bibnamefont{Winter}},
  \bibinfo{journal}{Phys. Rev. D}
  \textbf{\bibinfo{volume}{96}}(\bibinfo{number}{8}), \bibinfo{pages}{083018}
  (\bibinfo{year}{2017}), \eprint{1707.07684}.

\bibitem{Coloma:2018idr}
\bibinfo{author}{\bibfnamefont{P.}~\bibnamefont{Coloma}},
  \bibinfo{author}{\bibfnamefont{J.}~\bibnamefont{Lopez-Pavon}},
  \bibinfo{author}{\bibfnamefont{I.}~\bibnamefont{Martinez-Soler}},
  \bibnamefont{and} \bibinfo{author}{\bibfnamefont{H.}~\bibnamefont{Nunokawa}},
  \bibinfo{journal}{Eur. Phys. J.}
  \textbf{\bibinfo{volume}{C78}}(\bibinfo{number}{8}), \bibinfo{pages}{614}
  (\bibinfo{year}{2018}), \eprint{1803.04438}.

\bibitem{Gomes:2018inp}
\bibinfo{author}{\bibfnamefont{G.~B.} \bibnamefont{Gomes}},
  \bibinfo{author}{\bibfnamefont{D.~V.} \bibnamefont{Forero}},
  \bibinfo{author}{\bibfnamefont{M.~M.} \bibnamefont{Guzzo}},
  \bibinfo{author}{\bibfnamefont{P.~C.} \bibnamefont{de~Holanda}},
  \bibnamefont{and} \bibinfo{author}{\bibfnamefont{R.~L.~N.}
  \bibnamefont{Oliveira}}, \bibinfo{journal}{Phys. Rev. D}
  \textbf{\bibinfo{volume}{100}}, \bibinfo{pages}{055023}
  (\bibinfo{year}{2019}).

\bibitem{Carpio:2018gum}
\bibinfo{author}{\bibfnamefont{J.~A.} \bibnamefont{Carpio}},
  \bibinfo{author}{\bibfnamefont{E.}~\bibnamefont{Massoni}}, \bibnamefont{and}
  \bibinfo{author}{\bibfnamefont{A.~M.} \bibnamefont{Gago}},
  \bibinfo{journal}{Phys. Rev. D}
  \textbf{\bibinfo{volume}{100}}(\bibinfo{number}{1}), \bibinfo{pages}{015035}
  (\bibinfo{year}{2019}), \eprint{1811.07923}.

\bibitem{deGouvea:2020hfl}
\bibinfo{author}{\bibfnamefont{A.}~\bibnamefont{de~Gouvea}},
  \bibinfo{author}{\bibfnamefont{V.}~\bibnamefont{de~Romeri}},
  \bibnamefont{and} \bibinfo{author}{\bibfnamefont{C.~A.}
  \bibnamefont{Ternes}}, \bibinfo{journal}{JHEP} \textbf{\bibinfo{volume}{08}},
  \bibinfo{pages}{018} (\bibinfo{year}{2020}), \eprint{2005.03022}.

\bibitem{Gomes:2020muc}
\bibinfo{author}{\bibfnamefont{A.~L.~G.} \bibnamefont{Gomes}},
  \bibinfo{author}{\bibfnamefont{R.~A.} \bibnamefont{Gomes}}, \bibnamefont{and}
  \bibinfo{author}{\bibfnamefont{O.~L.~G.} \bibnamefont{Peres}}
  \eprint{arXiv:2001.09250 [hep-ph]}.

\bibitem{Nieves:2020jjg}
\bibinfo{author}{\bibfnamefont{J.~F.} \bibnamefont{Nieves}} \bibnamefont{and}
  \bibinfo{author}{\bibfnamefont{S.}~\bibnamefont{Sahu}},
  \bibinfo{journal}{Phys. Rev. D}
  \textbf{\bibinfo{volume}{102}}(\bibinfo{number}{5}), \bibinfo{pages}{056007}
  (\bibinfo{year}{2020}), \eprint{2002.08315}.

\bibitem{Stuttard:2020qfv}
\bibinfo{author}{\bibfnamefont{T.}~\bibnamefont{Stuttard}} \bibnamefont{and}
  \bibinfo{author}{\bibfnamefont{M.}~\bibnamefont{Jensen}},
  \bibinfo{journal}{Phys. Rev. D}
  \textbf{\bibinfo{volume}{102}}(\bibinfo{number}{11}), \bibinfo{pages}{115003}
  (\bibinfo{year}{2020}), \eprint{2007.00068}.

\bibitem{Porto-Silva:2021ael}
\bibinfo{author}{\bibfnamefont{Y.~P.} \bibnamefont{Porto-Silva}}
  \bibnamefont{and} \bibinfo{author}{\bibfnamefont{A.~Y.}
  \bibnamefont{Smirnov}}, \bibinfo{journal}{JCAP}
  \textbf{\bibinfo{volume}{06}}, \bibinfo{pages}{029} (\bibinfo{year}{2021}),
  \eprint{2103.10149}.

\bibitem{epr}
\bibinfo{author}{\bibfnamefont{A.}~\bibnamefont{Einstein}},
  \bibinfo{author}{\bibfnamefont{B.}~\bibnamefont{Podolsky}}, \bibnamefont{and}
  \bibinfo{author}{\bibfnamefont{N.}~\bibnamefont{Rosen}},
  \bibinfo{journal}{Phys. Rev.} \textbf{\bibinfo{volume}{47}},
  \bibinfo{pages}{777} (\bibinfo{year}{1935}),
  \urlprefix\url{https://link.aps.org/doi/10.1103/PhysRev.47.777}.

\bibitem{Bell:1964fg}
\bibinfo{author}{\bibfnamefont{J.~S.} \bibnamefont{Bell}},
  \bibinfo{journal}{Rev. Mod. Phys.} \textbf{\bibinfo{volume}{38}},
  \bibinfo{pages}{447} (\bibinfo{year}{1966}).

\bibitem{speakable}
\bibinfo{author}{\bibfnamefont{J.~S.} \bibnamefont{Bell}},
  \emph{\bibinfo{title}{Speakable and Unspeakable in Quantum Mechanics}}
  (\bibinfo{publisher}{Cambridge University Press, UK}, \bibinfo{year}{1987}).

\bibitem{Leggett:1985zz}
\bibinfo{author}{\bibfnamefont{A.~J.} \bibnamefont{Leggett}} \bibnamefont{and}
  \bibinfo{author}{\bibfnamefont{A.}~\bibnamefont{Garg}},
  \bibinfo{journal}{Phys. Rev. Lett.} \textbf{\bibinfo{volume}{54}},
  \bibinfo{pages}{857} (\bibinfo{year}{1985}).

\bibitem{emary2013leggett}
\bibinfo{author}{\bibfnamefont{C.}~\bibnamefont{Emary}},
  \bibinfo{author}{\bibfnamefont{N.}~\bibnamefont{Lambert}}, \bibnamefont{and}
  \bibinfo{author}{\bibfnamefont{F.}~\bibnamefont{Nori}},
  \bibinfo{journal}{Reports on Progress in Physics}
  \textbf{\bibinfo{volume}{77}}(\bibinfo{number}{1}), \bibinfo{pages}{016001}
  (\bibinfo{year}{2013}), ISSN \bibinfo{issn}{1361-6633},
  \urlprefix\url{http://dx.doi.org/10.1088/0034-4885/77/1/016001}.

\bibitem{ipsika2021}
\bibinfo{author}{\bibfnamefont{A.~V.} \bibnamefont{Varma}},
  \bibinfo{author}{\bibfnamefont{I.}~\bibnamefont{Mohanty}}, \bibnamefont{and}
  \bibinfo{author}{\bibfnamefont{S.}~\bibnamefont{Das}},
  \bibinfo{journal}{Journal of Physics A: Mathematical and Theoretical}
  \textbf{\bibinfo{volume}{54}}(\bibinfo{number}{11}), \bibinfo{pages}{115301}
  (\bibinfo{year}{2021}), ISSN \bibinfo{issn}{1751-8121},
  \urlprefix\url{http://dx.doi.org/10.1088/1751-8121/abde76}.

\bibitem{Karthik:2019wbp}
\bibinfo{author}{\bibfnamefont{H.~S.} \bibnamefont{Karthik}},
  \bibinfo{author}{\bibfnamefont{A.}~\bibnamefont{Shenoy~Hejamadi}},
  \bibnamefont{and} \bibinfo{author}{\bibfnamefont{A.~R.~U.}
  \bibnamefont{Devi}}, \bibinfo{journal}{Phys. Rev. A}
  \textbf{\bibinfo{volume}{103}}(\bibinfo{number}{3}), \bibinfo{pages}{032420}
  (\bibinfo{year}{2021}), \eprint{1907.02879}.

\bibitem{Chanda:2018vmf}
\bibinfo{author}{\bibfnamefont{T.}~\bibnamefont{Chanda}},
  \bibinfo{author}{\bibfnamefont{T.}~\bibnamefont{Das}},
  \bibinfo{author}{\bibfnamefont{S.}~\bibnamefont{Mal}},
  \bibinfo{author}{\bibfnamefont{A.~S.} \bibnamefont{De}}, \bibnamefont{and}
  \bibinfo{author}{\bibfnamefont{U.}~\bibnamefont{Sen}},
  \bibinfo{journal}{Phys. Rev. A}
  \textbf{\bibinfo{volume}{98}}(\bibinfo{number}{2}), \bibinfo{pages}{022138}
  (\bibinfo{year}{2018}), \eprint{1805.04665}.

\bibitem{Ghosh:2021ncf}
\bibinfo{author}{\bibfnamefont{S.}~\bibnamefont{Ghosh}},
  \bibinfo{author}{\bibfnamefont{A.~V.} \bibnamefont{Varma}}, \bibnamefont{and}
  \bibinfo{author}{\bibfnamefont{S.}~\bibnamefont{Das}}
  \eprint{arXiv:2110.10696 [quant-ph]}.

\bibitem{Zhang:2020dva}
\bibinfo{author}{\bibfnamefont{W.}~\bibnamefont{Zhang}},
  \bibinfo{author}{\bibfnamefont{R.~K.} \bibnamefont{Saripalli}},
  \bibinfo{author}{\bibfnamefont{J.~M.} \bibnamefont{Leamer}},
  \bibinfo{author}{\bibfnamefont{R.~T.} \bibnamefont{Glasser}},
  \bibnamefont{and} \bibinfo{author}{\bibfnamefont{D.~I.}
  \bibnamefont{Bondar}}, \bibinfo{journal}{Phys. Rev. A}
  \textbf{\bibinfo{volume}{104}}(\bibinfo{number}{4}), \bibinfo{pages}{043711}
  (\bibinfo{year}{2021}), \eprint{2012.03940}.

\bibitem{Formaggio:2016cuh}
\bibinfo{author}{\bibfnamefont{J.~A.} \bibnamefont{Formaggio}},
  \bibinfo{author}{\bibfnamefont{D.~I.} \bibnamefont{Kaiser}},
  \bibinfo{author}{\bibfnamefont{M.~M.} \bibnamefont{Murskyj}},
  \bibnamefont{and} \bibinfo{author}{\bibfnamefont{T.~E.} \bibnamefont{Weiss}},
  \bibinfo{journal}{Phys. Rev. Lett.}
  \textbf{\bibinfo{volume}{117}}(\bibinfo{number}{5}), \bibinfo{pages}{050402}
  (\bibinfo{year}{2016}), \eprint{1602.00041}.

\bibitem{Fu:2017hky}
\bibinfo{author}{\bibfnamefont{Q.}~\bibnamefont{Fu}} \bibnamefont{and}
  \bibinfo{author}{\bibfnamefont{X.}~\bibnamefont{Chen}},
  \bibinfo{journal}{Eur. Phys. J.}
  \textbf{\bibinfo{volume}{C77}}(\bibinfo{number}{11}), \bibinfo{pages}{775}
  (\bibinfo{year}{2017}).

\bibitem{Gangopadhyay:2013aha}
\bibinfo{author}{\bibfnamefont{D.}~\bibnamefont{Gangopadhyay}},
  \bibinfo{author}{\bibfnamefont{D.}~\bibnamefont{Home}}, \bibnamefont{and}
  \bibinfo{author}{\bibfnamefont{A.~S.} \bibnamefont{Roy}},
  \bibinfo{journal}{Phys. Rev.}
  \textbf{\bibinfo{volume}{A88}}(\bibinfo{number}{2}), \bibinfo{pages}{022115}
  (\bibinfo{year}{2013}).

\bibitem{Gangopadhyay:2017nsn}
\bibinfo{author}{\bibfnamefont{D.}~\bibnamefont{Gangopadhyay}}
  \bibnamefont{and} \bibinfo{author}{\bibfnamefont{A.~S.} \bibnamefont{Roy}},
  \bibinfo{journal}{Eur. Phys. J.}
  \textbf{\bibinfo{volume}{C77}}(\bibinfo{number}{4}), \bibinfo{pages}{260}
  (\bibinfo{year}{2017}).

\bibitem{Naikoo:2017fos}
\bibinfo{author}{\bibfnamefont{J.}~\bibnamefont{Naikoo}},
  \bibinfo{author}{\bibfnamefont{A.~K.} \bibnamefont{Alok}},
  \bibinfo{author}{\bibfnamefont{S.}~\bibnamefont{Banerjee}},
  \bibinfo{author}{\bibfnamefont{S.}~\bibnamefont{Uma~Sankar}},
  \bibinfo{author}{\bibfnamefont{G.}~\bibnamefont{Guarnieri}},
  \bibinfo{author}{\bibfnamefont{C.}~\bibnamefont{Schultze}}, \bibnamefont{and}
  \bibinfo{author}{\bibfnamefont{B.~C.} \bibnamefont{Hiesmayr}},
  \bibinfo{journal}{Nucl. Phys. B} \textbf{\bibinfo{volume}{951}},
  \bibinfo{pages}{114872} (\bibinfo{year}{2020}).

\bibitem{Richter:2017toa}
\bibinfo{author}{\bibfnamefont{M.}~\bibnamefont{Richter}},
  \bibinfo{author}{\bibfnamefont{B.}~\bibnamefont{Dziewit}}, \bibnamefont{and}
  \bibinfo{author}{\bibfnamefont{J.}~\bibnamefont{Dajka}},
  \bibinfo{journal}{Phys. Rev. D}
  \textbf{\bibinfo{volume}{96}}(\bibinfo{number}{7}), \bibinfo{pages}{076008}
  (\bibinfo{year}{2017}).

\bibitem{Naikoo:2018amb}
\bibinfo{author}{\bibfnamefont{J.}~\bibnamefont{Naikoo}} \bibnamefont{and}
  \bibinfo{author}{\bibfnamefont{S.}~\bibnamefont{Banerjee}},
  \bibinfo{journal}{Eur. Phys. J. C}
  \textbf{\bibinfo{volume}{78}}(\bibinfo{number}{7}), \bibinfo{pages}{602}
  (\bibinfo{year}{2018}), \eprint{1808.00365}.

\bibitem{Naikoo:2019eec}
\bibinfo{author}{\bibfnamefont{J.}~\bibnamefont{Naikoo}},
  \bibinfo{author}{\bibfnamefont{A.~K.} \bibnamefont{Alok}},
  \bibinfo{author}{\bibfnamefont{S.}~\bibnamefont{Banerjee}}, \bibnamefont{and}
  \bibinfo{author}{\bibfnamefont{S.~U.} \bibnamefont{Sankar}},
  \bibinfo{journal}{Phys. Rev. D} \textbf{\bibinfo{volume}{99}},
  \bibinfo{pages}{095001} (\bibinfo{year}{2019}).

\bibitem{Alok:2014gya}
\bibinfo{author}{\bibfnamefont{A.~K.} \bibnamefont{Alok}},
  \bibinfo{author}{\bibfnamefont{S.}~\bibnamefont{Banerjee}}, \bibnamefont{and}
  \bibinfo{author}{\bibfnamefont{S.~U.} \bibnamefont{Sankar}},
  \bibinfo{journal}{Nucl. Phys.} \textbf{\bibinfo{volume}{B909}},
  \bibinfo{pages}{65} (\bibinfo{year}{2016}).

\bibitem{Dixit:2018kev}
\bibinfo{author}{\bibfnamefont{K.}~\bibnamefont{Dixit}},
  \bibinfo{author}{\bibfnamefont{J.}~\bibnamefont{Naikoo}},
  \bibinfo{author}{\bibfnamefont{S.}~\bibnamefont{Banerjee}}, \bibnamefont{and}
  \bibinfo{author}{\bibfnamefont{A.~K.} \bibnamefont{Alok}},
  \bibinfo{journal}{Eur. Phys. J.}
  \textbf{\bibinfo{volume}{C78}}(\bibinfo{number}{11}), \bibinfo{pages}{914}
  (\bibinfo{year}{2018}).

\bibitem{Dixit:2018gjc}
\bibinfo{author}{\bibfnamefont{K.}~\bibnamefont{Dixit}},
  \bibinfo{author}{\bibfnamefont{J.}~\bibnamefont{Naikoo}},
  \bibinfo{author}{\bibfnamefont{S.}~\bibnamefont{Banerjee}}, \bibnamefont{and}
  \bibinfo{author}{\bibfnamefont{A.}~\bibnamefont{Kumar~Alok}},
  \bibinfo{journal}{Eur. Phys. J. C}
  \textbf{\bibinfo{volume}{79}}(\bibinfo{number}{2}), \bibinfo{pages}{96}
  (\bibinfo{year}{2019}), \eprint{1809.09947}.

\bibitem{Dixit:2019swl}
\bibinfo{author}{\bibfnamefont{K.}~\bibnamefont{Dixit}} \bibnamefont{and}
  \bibinfo{author}{\bibfnamefont{A.}~\bibnamefont{Kumar~Alok}},
  \bibinfo{journal}{Eur. Phys. J. Plus}
  \textbf{\bibinfo{volume}{136}}(\bibinfo{number}{3}), \bibinfo{pages}{334}
  (\bibinfo{year}{2021}), \eprint{1909.04887}.

\bibitem{Song:2018bma}
\bibinfo{author}{\bibfnamefont{X.-K.} \bibnamefont{Song}},
  \bibinfo{author}{\bibfnamefont{Y.}~\bibnamefont{Huang}},
  \bibinfo{author}{\bibfnamefont{J.}~\bibnamefont{Ling}}, \bibnamefont{and}
  \bibinfo{author}{\bibfnamefont{M.-H.} \bibnamefont{Yung}},
  \bibinfo{journal}{Phys. Rev. A}
  \textbf{\bibinfo{volume}{98}}(\bibinfo{number}{5}), \bibinfo{pages}{050302}
  (\bibinfo{year}{2018}).

\bibitem{Ming:2020nyc}
\bibinfo{author}{\bibfnamefont{F.}~\bibnamefont{Ming}},
  \bibinfo{author}{\bibfnamefont{X.-K.} \bibnamefont{Song}},
  \bibinfo{author}{\bibfnamefont{J.}~\bibnamefont{Ling}},
  \bibinfo{author}{\bibfnamefont{L.}~\bibnamefont{Ye}}, \bibnamefont{and}
  \bibinfo{author}{\bibfnamefont{D.}~\bibnamefont{Wang}},
  \bibinfo{journal}{Eur. Phys. J. C}
  \textbf{\bibinfo{volume}{80}}(\bibinfo{number}{3}), \bibinfo{pages}{275}
  (\bibinfo{year}{2020}).

\bibitem{Blasone:2021cau}
\bibinfo{author}{\bibfnamefont{M.}~\bibnamefont{Blasone}},
  \bibinfo{author}{\bibfnamefont{S.}~\bibnamefont{De~Siena}}, \bibnamefont{and}
  \bibinfo{author}{\bibfnamefont{C.}~\bibnamefont{Matrella}}
  \eprint{arXiv:2104.03166 [quant-ph]}.

\bibitem{Shafaq:2020sqo}
\bibinfo{author}{\bibfnamefont{S.}~\bibnamefont{Shafaq}} \bibnamefont{and}
  \bibinfo{author}{\bibfnamefont{P.}~\bibnamefont{Mehta}}, \bibinfo{journal}{J.
  Phys. G} \textbf{\bibinfo{volume}{48}}(\bibinfo{number}{8}),
  \bibinfo{pages}{085002} (\bibinfo{year}{2021}), \eprint{2009.12328}.

\bibitem{Sarkar:2020vob}
\bibinfo{author}{\bibfnamefont{T.}~\bibnamefont{Sarkar}} \bibnamefont{and}
  \bibinfo{author}{\bibfnamefont{K.}~\bibnamefont{Dixit}},
  \bibinfo{journal}{Eur. Phys. J. C}
  \textbf{\bibinfo{volume}{81}}(\bibinfo{number}{1}), \bibinfo{pages}{88}
  (\bibinfo{year}{2021}), \eprint{2010.02175}.

\bibitem{Mehta:2009ea}
\bibinfo{author}{\bibfnamefont{P.}~\bibnamefont{Mehta}},
  \bibinfo{journal}{Phys. Rev. D} \textbf{\bibinfo{volume}{79}},
  \bibinfo{pages}{096013} (\bibinfo{year}{2009}).

\bibitem{mns}
\bibinfo{author}{\bibfnamefont{Z.}~\bibnamefont{Maki}},
  \bibinfo{author}{\bibfnamefont{M.}~\bibnamefont{Nakagawa}}, \bibnamefont{and}
  \bibinfo{author}{\bibfnamefont{S.}~\bibnamefont{Sakata}},
  \bibinfo{journal}{Prog. Theo. Phys.}
  \textbf{\bibinfo{volume}{28}}(\bibinfo{number}{5}), \bibinfo{pages}{870}
  (\bibinfo{year}{1962}).

\bibitem{Gribov:1968kq}
\bibinfo{author}{\bibfnamefont{V.}~\bibnamefont{Gribov}} \bibnamefont{and}
  \bibinfo{author}{\bibfnamefont{B.}~\bibnamefont{Pontecorvo}},
  \bibinfo{journal}{Phys. Lett. B} \textbf{\bibinfo{volume}{28}},
  \bibinfo{pages}{493} (\bibinfo{year}{1969}).

\bibitem{Jarlskog:1985ht}
\bibinfo{author}{\bibfnamefont{C.}~\bibnamefont{Jarlskog}},
  \bibinfo{journal}{Phys. Rev. Lett.} \textbf{\bibinfo{volume}{55}},
  \bibinfo{pages}{1039} (\bibinfo{year}{1985}).

\bibitem{Lindblad:1975ef}
\bibinfo{author}{\bibfnamefont{G.}~\bibnamefont{Lindblad}},
  \bibinfo{journal}{Commun. Math. Phys.} \textbf{\bibinfo{volume}{48}},
  \bibinfo{pages}{119} (\bibinfo{year}{1976}).

\bibitem{Ellis1984381}
\bibinfo{author}{\bibfnamefont{J.}~\bibnamefont{Ellis}},
  \bibinfo{author}{\bibfnamefont{J.~S.} \bibnamefont{Hagelin}},
  \bibinfo{author}{\bibfnamefont{D.}~\bibnamefont{Nanopoulos}},
  \bibnamefont{and}
  \bibinfo{author}{\bibfnamefont{M.}~\bibnamefont{Srednicki}},
  \bibinfo{journal}{Nuclear Physics B}
  \textbf{\bibinfo{volume}{241}}(\bibinfo{number}{2}), \bibinfo{pages}{381 }
  (\bibinfo{year}{1984}), ISSN \bibinfo{issn}{0550-3213}.

\bibitem{Banks:1983by}
\bibinfo{author}{\bibfnamefont{T.}~\bibnamefont{Banks}},
  \bibinfo{author}{\bibfnamefont{L.}~\bibnamefont{Susskind}}, \bibnamefont{and}
  \bibinfo{author}{\bibfnamefont{M.~E.} \bibnamefont{Peskin}},
  \bibinfo{journal}{Nucl. Phys. B} \textbf{\bibinfo{volume}{244}},
  \bibinfo{pages}{125} (\bibinfo{year}{1984}).

\bibitem{Liu:1993ji}
\bibinfo{author}{\bibfnamefont{J.}~\bibnamefont{Liu}},
  \bibinfo{journal}{Physics Letters B}
  \textbf{\bibinfo{volume}{314}}(\bibinfo{number}{1}), \bibinfo{pages}{52}
  (\bibinfo{year}{1993}), ISSN \bibinfo{issn}{0370-2693},
  \urlprefix\url{https://www.sciencedirect.com/science/article/pii/037026939391320M}.

\bibitem{Benatti:2000ph}
\bibinfo{author}{\bibfnamefont{F.}~\bibnamefont{Benatti}} \bibnamefont{and}
  \bibinfo{author}{\bibfnamefont{R.}~\bibnamefont{Floreanini}},
  \bibinfo{journal}{JHEP} \textbf{\bibinfo{volume}{02}}, \bibinfo{pages}{032}
  (\bibinfo{year}{2000}), \eprint{hep-ph/0002221}.

\bibitem{Mavromatos:2007hv}
\bibinfo{author}{\bibfnamefont{N.~E.} \bibnamefont{Mavromatos}},
  \bibinfo{author}{\bibfnamefont{A.}~\bibnamefont{Meregaglia}},
  \bibinfo{author}{\bibfnamefont{A.}~\bibnamefont{Rubbia}},
  \bibinfo{author}{\bibfnamefont{A.}~\bibnamefont{Sakharov}}, \bibnamefont{and}
  \bibinfo{author}{\bibfnamefont{S.}~\bibnamefont{Sarkar}},
  \bibinfo{journal}{Phys. Rev. D} \textbf{\bibinfo{volume}{77}},
  \bibinfo{pages}{053014} (\bibinfo{year}{2008}), \eprint{0801.0872}.

\end{thebibliography}

\end{document}